\pdfoutput=1
\DocumentMetadata{} % hyperref error fix
\documentclass[sigconf]{acmart}
\usepackage{xkeyval}

\AtBeginDocument{%
  \providecommand\BibTeX{{%
    \normalfont B\kern-0.5em{\scshape i\kern-0.25em b}\kern-0.8em\TeX}}}

\copyrightyear{2026}
\acmYear{2026}
\setcopyright{cc}
\setcctype{by}
\acmConference[WSDM '26]{Proceedings of the Nineteenth ACM International Conference on Web Search and Data Mining}{February 22--26, 2026}{Boise, ID, USA}
\acmBooktitle{Proceedings of the Nineteenth ACM International Conference on Web Search and Data Mining (WSDM '26), February 22--26, 2026, Boise, ID, USA}
\acmPrice{}
\acmDOI{10.1145/3773966.3777977}
\acmISBN{979-8-4007-2292-9/2026/02}

\usepackage[normalem]{ulem}
\useunder{\uline}{\ul}{}
\usepackage{placeins} % 문서의 preamble에 추가

\usepackage{balance}
\usepackage{pifont}
\usepackage{amsfonts, amsthm, amsmath}
\usepackage[ruled,vlined,linesnumbered]{algorithm2e}
\usepackage{color, colortbl}
\usepackage{enumitem,multirow,graphicx,subcaption,multicol,lipsum,float,adjustbox}
\usepackage{hyperref}
\usepackage{hyperxmp}
\newlength{\textfloatsepsave}  
\usepackage{cleveref}
\usepackage{algorithmicx}
\usepackage[page]{appendix} % print appendices title
\crefformat{section}{\S#2#1#3}
\crefformat{subsection}{\S#2#1#3}
\crefformat{subsubsection}{\S#2#1#3}
\begin{document}


\title{Capturing User Interests from Data Streams \\for Continual Sequential Recommendation}
% Collaborative Knowledge Distillation for Continual Learning in Recommender System
% Continual Collaborative Distillation for Recommender System

\begin{abstract}
Transformer-based sequential recommendation (SR) models excel at modeling long-range dependencies, but suffer from high computational costs and catastrophic forgetting during continuous updates.
Although continual learning has been applied to recommendation, existing methods gradually forget long-term user preferences and remain underexplored in SR.
In this paper, we introduce
\textbf{\underline{C}}ontinual 
\textbf{\underline{S}}equential
\textbf{\underline{T}}ransformer for \textbf{\underline{Rec}}ommendation (\textbf{\proposed}), which effectively adapt to current interests by leveraging preserved historical knowledge.
Its core is \textbf{Continual Sequential Attention (CSA)}, a linear attention tailored for continual SR, which partially retain historical knowledge without direct access to prior data. 
% To this end, we introduce \textit{Continual Sequential Attention (CSA)}, which tailors linear attention for continual SR, enabling the effective retention of historical knowledge without direct access to all historical data. 
CSA features: (1) \textit{Cauchy-Schwarz Normalization} to stabilize learning over time under uneven user interaction frequencies, and (2) \textit{Collaborative Interest Enrichment} via shared, learnable interest pools to mitigate forgetting.
We also introduce a new technique for new user adaptation by transferring historical knowledge from existing users with similar interests.
% \textit{Pseudo-Historical Knowledge Assignment} to facilitate the adaptation of new users by leveraging the historical knowledge of existing users with similar behavioral patterns.
% Additionally, we present \textit{Pseudo-Historical Knowledge Assignment} to facilitate the adaptation of new users by leveraging the historical knowledge of existing users with similar behavioral patterns. %, thereby effectively integrating new users into the CSA computation.
Extensive experiments show \proposed's superior performance in both knowledge retention and acquisition.\footnote{Our code is available at \href{https://github.com/Gyu-Seok0/CSTRec_WSDM26}{https://github.com/Gyu-Seok0/CSTRec\_WSDM26}.}
% We expect that \proposed can significantly enhance the practical deployment of SR in real-world applications by addressing challenges posed by non-stationary data streams and continuously evolving user behaviors.
%Our code is available at \href{https://github.com/Gyu-Seok0/CSTRec_WSDM26}{\color{magenta}{https://github.com/Gyu-Seok0/CSTRec\_WSDM26}}.
\vspace{-0.5cm}

% Generated Abstract

\end{abstract}

\vspace{-1cm}


\vskip -10em % <--- 강력한 TeX 명령어 사용 (작은 음수 값부터 테스트)
\addvspace{-10pt} % <--- LaTeX 환경의 간격을 무시하도록 시도
% methods
\newcommand{\proposed}{CSTRec\xspace}
\newcommand{\tRS}{$RS^T$\xspace}
\newcommand{\sRS}{$RS^S$\xspace}

\newcommand{\smallsection}[1]{{\vspace{0.03in} \noindent \bf {#1}}}

\author{Gyuseok Lee}
\affiliation{
    \institution{University of Illinois Urbana-Champaign	}
    \city{Champaign}
    \state{IL}
    \country{USA}
}
\email{gyuseok2@illinois.edu}

\author{Hyunsik Yoo}
\affiliation{
    \institution{University of Illinois Urbana-Champaign	}
    \city{Champaign}
    \state{IL}
    \country{USA}
}
\email{hy40@illinois.edu}

\author{Junyoung Hwang}
\affiliation{
    \institution{42dot}
    \city{Seongnam}
    %\state{Gyeongbuk}
    \country{Republic of Korea}
}
\email{junyoung.hwang@42dot.ai}

\author{SeongKu Kang}
\affiliation{
    \city{}
    \institution{Korea University}
    \city{Seoul}
    %\state{IL}
    \country{Republic of Korea}
}
\authornote{Corresponding author.}
\email{seongkukang@korea.ac.kr}

\author{Hwanjo Yu}
\affiliation{
    \institution{Pohang University of \\ Science and Technology}
    \city{Pohang}
    %\state{Gyeongbuk\\}
    \country{Republic of Korea}
}
\email{hwanjoyu@postech.ac.kr}

% \author{Craig Rodkin}
% \affiliation{%
%    \institution{ACM}
%    \city{New York City}
%    \state{NY}
%    \country{USA}}
% \email{rodkin@hq.acm.org}

\maketitle
\vspace{-0.5cm}
\section{Introduction}
Sequential recommendation (SR) has gained prominence in both academia and practical applications by capturing sequential patterns in user behavior to enhance next item prediction~\cite{sasrec,fang2020deep,steck2021deep,yue2024linear}.
Earlier studies use Markov chains~\cite{FPMC, chen2012playlist} and Recurrent Neural Networks (RNNs)~\cite{wu2017recurrent, gru4rec} to model temporal dependencies in user behavior sequences.
However, both approaches estruggle to capture long-range dependencies as sequence length increases~\cite{sasrec}.
To address this, Transformer-based SR models~\cite{sasrec,bert4rec,s3rec, personalized_transformer, STOSA} employ self-attention to focus on the most relevant behaviors regardless of position~\cite{transformer}.
With its remarkable performance, Transformer has become the dominant~approach~in~SR.

Nevertheless, applying Transformer-based SR models to real-world applications presents non-trivial challenges.
As user behavior sequences are continuously arriving, models should adapt to new information for timely recommendations~\cite{cho2021learning, li2020time,costa2019collective}.
A simple approach is to retrain the model using the entire user behavior sequences accumulated along data streams.
However, training on the whole sequences is extremely time-consuming~\cite{IMSR, mi2020ader}, due to the quadratic complexity of self-attention with respect to input length~\cite{linear_attention,linformer,performer}.
Moreover, this approach is impractical in scenarios where not all historical interactions are accessible (e.g., privacy issues) or where memory constraints prevent storing complete sequences during both training and inference~\cite{zhang2023survey, IMSR}.
A more cost-effective alternative is to fine-tune the model using only new sequences.
However, such overreliance on transient interests risks forgetting historical user interests that may reemerge later, thereby significantly limiting recommendation accuracy~\cite{IMSR, GraphSAIL, PIW}.

Continual learning (CL)~\cite{lee2024continual, kirkpatrick2017overcoming, li2017learning}, a well-established approach to updating a model with non-stationary data streams, has been actively studied for recommendation~\cite{GraphSAIL, mi2020ader, LWCKD, PIW, zhu2023reloop2}. 
With the goal of adapting to new data while preserving previously learned knowledge, there are two dominant CL approaches: (1) Regularization-based methods~\cite{LWCKD,GraphSAIL,PIW} impose constraints on the parameter space to prevent significant changes from previously trained parameters.
(2) Replay-based methods~\cite{mi2020ader, cai2022reloop, zhu2023reloop2} store small portions of historical data and reuse them in subsequent training, with external memory being updated over time.

Although existing CL methods retain interests from the \textit{relatively recent} past, they are insufficient to preserve those from the \textit{distant} past, gradually forgetting long‑term preferences.
Specifically, regularization-based methods rely on the most recently learned parameters to transfer acquired knowledge into the subsequent training process. 
As this greedy preservation continues, new information gradually dilutes historical knowledge, leading to unavoidable forgetting.
Similarly, replay-based methods are constrained by limited memory capacity, which necessitates continually updating stored data with newer interactions.
This reduces access to older data, further hindering the retention of historical user interests.
Furthermore, most CL methods have focused on non-sequential models (e.g., matrix factorization~\cite{jakomin2020simultaneous,yu2016incremental}, graph neural networks~\cite{LWCKD, PIW}), leaving their application in SR relatively underexplored. 

We propose \textbf{\underline{C}}ontinual \textbf{\underline{S}}equential \textbf{\underline{T}}ransformer for \textbf{\underline{Rec}}ommenda tion (\textbf{\proposed}), which continuously updates a transformer-based SR model with non-stationary data streams. 
\proposed aims to preserve historical interests and leverage them to adapt to current ones, ultimately capturing the trajectory of user interests over time.
To facilitate the preservation of historical knowledge, we borrow the idea of linear attention~\cite{linear_attention, linformer, performer}.
Linear attention approximates self-attention by performing linear computations at each position and sequentially accumulating hidden states over time, similar to the hidden state propagation in RNNs.
This allows the model to partially retain historical knowledge through parametric memories without direct access to all historical sequences, while emulating the expressive power of self‑attention~\cite{Infini-attention,lee2023recasting}.

However, naively applying linear attention to continual SR yields suboptimal results due to two challenges: 
(1) unstable training, caused by an imbalance in the number of interactions per user along data streams;
hidden states accumulate user behaviors over time, resulting in disproportionately large values for active users and small values for less active users.
This imbalance leads to uneven magnitudes and updates across users, making optimization highly unstable.
(2) inevitable forgetting, caused by the continual update of hidden states with new sequences over time. 
This causes historical knowledge to be continuously overwritten~and~gradually~forgotten.

As a solution, we introduce \textbf{Continual Sequential Attention (CSA)}, a specialized linear attention for continual SR, featuring two novel components:
(1) \textit{Cauchy-Schwarz Normalization} to resolve unstable learning.
Leveraging the Cauchy-Schwarz inequality, we dynamically adjust the magnitudes of hidden states to address the imbalance caused by disparities in the number of interactions.
(2) \textit{Collaborative Interest Enrichment} to alleviate inevitable forgetting.
We utilize learnable interest pools that store historical interests.
For each user context, we retrieve the most relevant interests from the pools and use them to compensate for forgotten user interests.
With CSA, \proposed not only inherits the strengths of linear attention but also effectively addresses its limitations for continual SR.
Furthermore, to facilitate the accommodation of newly joined users, we introduce a new technique called \textit{Pseudo-Historical Knowledge Assignment}.
By leveraging the historical knowledge of existing users with similar behavioral patterns, it allows new users to be effectively integrated into CSA computation.

Our contributions are summarized as follows:
\vspace{-0.7cm}
\begin{itemize}[leftmargin=*] \vspace{-\topsep}
    \item We highlight the challenges of employing transformer-based SR models in non-stationary data streams, which have not been studied well in the previous literature.
    To the best of our knowledge, we are the first to address these challenges.
  
    \item We propose \proposed equipped with CSA—a linear attention mechanism tailored for continual SR—to effectively retain historical knowledge and acquire current one, thereby capturing the trajectory of user interests over time.

    \item We validate the effectiveness of \proposed through comprehensive experiments on real-world datasets and provide in-depth analyses to validate each proposed component.
\end{itemize}

\vspace{-0.2cm}

\vspace{-0.3cm}
\section{Related Work}
\label{sec:relatedwork}
% Various sequential modeling approaches have been applied, starting from Markov chains (MC)~\cite{FPMC, shani2005mdp, chen2012playlist}. 
% SR is a well-established research area, 

% 전반적으로
\noindent
\textbf{Sequential Recommendation (SR).}
% SR aims to capture sequential patterns within user behavior sequences. % for next item prediction.
Earlier SR studies \cite{FPMC, shani2005mdp, chen2012playlist} employ Markov chains (MC).
FPMC \cite{FPMC} bridges matrix factorization and MC for next-basket recommendation, while LME~\cite{chen2012playlist} applies metric learning to learn Markov embeddings. % for next-playlist prediction.
With the advent of deep learning, RNNs have been applied in SR~\cite{wu2017recurrent, gru4rec, tan2016improved, narm}.
GRU4Rec \cite{gru4rec} pioneers Gated Recurrent Units (GRUs), % for session-based recommendations.
and \cite{tan2016improved} further improves RNN-based models by augmenting data and addressing data distribution shifts. 
NARM \cite{narm} employs two different GRUs to encode both the sequential pattern and the main interest within a given session.
% Transformer
Recently, transformer-based SR models \cite{sasrec, bert4rec, s3rec, personalized_transformer, STOSA} have shown exceptional performance by effectively capturing long-range dependencies via self-attention, thereby becoming dominant in SR.
% The self-attention mechanism focuses on the most relevant behaviors, irrespective of their position within the sequence~\cite{transformer}.
SASRec~\cite{sasrec} pioneers the use of self-attention~\cite{transformer}, and BERT4Rec \cite{bert4rec} applies bidirectional self-attention~\cite{bert}.
SSE-PT~\cite{personalized_transformer} addresses the lack of personalization in the Transformer by employing stochastic shared embeddings~\cite{wu2019stochastic}.
% 문제점

However, updating Transformer-based SR models along data streams poses two non-trivial challenges: 
high training costs and catastrophic forgetting, which remain  underexplored. 
Further study is needed to extend the applicability of SR models to real-world applications with continuously arriving user behavior~sequences.

\smallsection{Continual Learning (CL).}
\label{sub:cl}
%Also known as lifelong learning or incremental learning, 
CL is a well-established research area to update a model along data streams~\cite{lee2024continual, yoo2025continualb, yoo2025continuala}.
% inspired by human lifelong learning~\cite{dualnet, arani2022learning, zenke2017continual}, 
The goal of CL is to effectively balance knowledge acquisition and retention over time~\cite{jung2023new, kim2023stability, yoo2024ensuring}.
% The goal of CL is to balance the \textit{stability-plasticity dilemma}: retaining previously learned knowledge (\textit{stability}) while rapidly adapting to new data (\textit{plasticity}) over time~\cite{jung2023new, kim2023stability}.
Recently, CL has been actively studied for recommendation \cite{GraphSAIL, mi2020ader, LWCKD, PIW, cai2022reloop, zhu2023reloop2, CL_RS_replay, he2023dynamically, yoo2025embracing} to rapidly adapt to new data while leveraging well-preserved historical knowledge.
Two key CL approaches are (1) regularization~\cite{kirkpatrick2017overcoming, li2017learning, LWCKD, PIW} and (2) experience replay~\cite{prabhu2020gdumb, rebuffi2017icarl, zhu2023reloop2, mi2020ader}.

Regularization-based methods penalize rapid changes to previously trained parameters.
% Xu et al.~\cite{GraphSAIL} introduce GraphSAIL, which applies knowledge distillation in GNNs at the node, neighborhood, and global levels.
For example, LWC-KD~\cite{LWCKD} introduces contrastive knowledge distillation between the parameters of the previously and currently trained graph neural networks.
SAIL-PIW~\cite{PIW} proposes personalized imitation weights to adjust knowledge retention based on user preferences being static or dynamic.
On the other hand, replay-based methods store and retrieve historical data from external memory, which is continuously updated.
ADER~\cite{mi2020ader} assigns memory slots to each item based on its frequency in data streams and retains (session, target item) pairs whose session features are closest to the average feature vector.
Reloop2~\cite{zhu2023reloop2} introduces a self-correcting loop that stores mispredicted samples in a non-parametric memory to improve future learning.

% 문제점
However, existing CL methods fall short of retaining long‑term user interests and thus suffer gradual forgetting.
This is primarily due to the sequential integration of previously learned knowledge into subsequent training processes.
As models continually adapt to non-stationary data streams, historical knowledge becomes diluted by new information and is gradually forgotten, making it difficult to capture long-term user preferences.
%\hs{I still do not fully understand this part. To me, "evolving/changing user preferences" sounds more like "short term preferences", or at least far from "long-range preferences". If that's the case, then gradually forgetting might actually help in capturing "short-term preferences or changing preferences". I feel it'd be clearer to use a term that conveys "long-range or stable knowledge" to make the connection to "gradually forgetting" more logical.}
Furthermore, most prior studies have focused solely on non-sequential models, leaving their application in SR relatively underexplored. 
Therefore, it is necessary to develop a  specialized approach tailored to continual SR.

\section{PRELIMINARIES}
\label{sec:preliminary}
\subsection{Problem Formulation}
We view the entire data stream $D$ as consecutive data blocks $[D_1, D_2, \\ \dots, D_t, \dots]$, where $D_t$ contains interaction sequences observed during the time period $t$ (e.g., weekly or monthly).
Let $\mathcal{U}^t$ and $\mathcal{I}^t$ be the set of users and items within $D_t$, respectively.
Let $S^t_u = [i^t_1, \dots, i^t_k, \dots, i^t_{|S^t_u|}]$ be the sequence of items with which user $u$ interacted in $D_t$, where $i^t_k$ denotes the $k$-th item in $S^t_u$. 
Traditional SR aims to predict the next item $i^t_{|S^t_u|+1}$ given  $S^t_u$.
Building upon this, our task (i.e., continual SR) is to predict the next item for each incoming interaction sequence (i.e., $S^1_u, S^2_u, \dots, S^t_u, \dots$).
Note that at each time period $t$, we update the model parameters solely on the newly arrived block $D_t$, without accessing any previous data $D_{<t}$.

% We define the lifelong sequence of user $u$ as $S_u = \{S^1_u, S^2_u, \dots, S^t_u, \dots \}$, which incrementally accumulates the user's behavior over time.
% In this work, we extend continual learning for recommender systems to lifelong sequential recommendation in a dynamic environment, where continuous stream of numerous and non-stationary user-item interactions arrives as incremental sequences (i.e., $S^1_u, \dots, S^t_u, \dots$).
% Our goal is to adapt to dynamic user interests within each incremental sequence (e.g., $S^t_u$) while preserving their static interests across the lifelong sequence (i.e., $S_u$), thereby capturing the evolving sequential patterns of user preferences over time. 
% Our framework operates under the constraint of not having direct access to historical data, ensuring practicality and scalability for deploying recommender systems in real-world applications.

% \vspace{-0.1cm}
\subsection{Background}
\subsubsection{\textbf{Transformer-based SR model.}}
\label{subsub:transformer}
% 이 섹션에서 뭘할려고 하는가 -> transformer 구조 설명.
The architecture of the Transformer-based SR models~\cite{sasrec} consists of multi-head attention (MH), a position-wise feed-forward network (FFN), layer normalization, and dropout. 
Given an input sequence $S^t_u = [i^t_1,\dots, i^t_N]$ of $N$ items, we embed each item into a $d$-dimensional vector and stack them to form $\mathbf{E}_{S^t_u} \in \mathbb{R}^{N \times d}$.
$\mathbf{E}_{S^t_u}$ serves as the initial hidden states $\mathbf{H}^0$, on which the self-attention operation is performed as follows:
% Since the introduction of self-attention in SR by SASRec~\cite{sasrec}, Transformer-based SR~\cite{transformer}, which builds upon self-attention, has become predominant approach \cite{bert4rec, s3rec, personalized_transformer, STOSA}.
% Self-attention allows recommenders to attend to different positions of items within an input sequence and represents their relationships as attention scores, effectively capturing long-term dependencies. 
% \vspace{-\topsep}
%\vspace{-0.1cm}
\begin{align}\label{eq:self_att}
    \text{head}_i &= \text{softmax}\bigg(\frac{(\mathbf{H}^{l-1}\mathbf{W}^{(i)}_Q)(\mathbf{H}^{l-1}\mathbf{W}^{(i)}_K)^\top}{\sqrt{d}} \bigg)(\mathbf{H}^{l-1}\mathbf{W}^{(i)}_V), \nonumber\\
    \text{MH}(\mathbf{H}^{l-1}) &= \text{Concat}(\text{head}_1, \dots, \text{head}_h) \mathbf{W}_O, \\
    \mathbf{G}^{l-1} &= \text{LayerNorm}(\mathbf{H}^{l-1} + \text{Dropout}(\text{MH}(\mathbf{H}^{l-1}))), \nonumber\\
    \mathbf{H}^{l} &= \text{LayerNorm}(\mathbf{G}^{l-1} + \text{Dropout}(\text{FFN}(\mathbf{G}^{l-1}))), \nonumber 
\end{align}
% \par\vspace{-0.1cm} % 줄바꿈
\noindent where $\mathbf{H}^l \in \mathbb{R}^{N \times d}$ is the hidden states at the $l$-th Transformer layer ($l = 1, \dots, L$).
The weights $\mathbf{W}^{(i)}_Q, \mathbf{W}^{(i)}_K, \mathbf{W}^{(i)}_V \in \mathbb{R}^{d \times (d/h)}$ are for the query, key, and value in attention $\text{head}_i$, where $h$ is the total number of heads.
$\mathbf{W}_O \in \mathbb{R}^{d \times d}$ projects the multi-head attention output.
We use $\mathbf{H}^L = [\mathbf{h}^L_1, \dots, \mathbf{h}^L_N]^\top$ (from the $L$-th Transformer layer) for next item prediction via a dot product with item embeddings.% during both training and inference as follows:
% \par\vspace{-0.04cm}

\smallsection{Training.}
We use the binary cross-entropy loss on block $D_t$:%\hs{at t?}
\small
\vspace{-\topsep}
\begin{equation}
\hspace{-0.4cm} % 기본 들여쓰기 제거
    \begin{aligned}
        \mathcal{L}_{\text{BCE}} = -\sum_{S^t_u \in D_t}\sum^{N-1}_{j=1}\Big[ \log \sigma  (\mathbf{e}^\top_{i^t_{j+1}}\mathbf{h}^L_j ) + \sum_{i_{\text{neg}} \in \mathcal{I}_{\text{neg}}}
        \log \big(1-\sigma  (\mathbf{e}^\top_{i_{\text{neg}}}\mathbf{h}^L_j )\big)
        \Big]
    \end{aligned}%\vspace{-0.1cm}
    \label{eq:bce}
\end{equation}
\normalsize
%\hs{does this loss also apply to $e_{i_{N+1}}$?  }
where $\sigma$ is the sigmoid function, $\mathcal{I}_{\text{neg}}\subset \mathcal{I}^{1:t} \setminus S_u^t$ is the set of randomly sampled negative items per time step and $\mathcal{I}^{1:t} = \bigcup^t_{t'=1}\mathcal{I}^{t'}$ is the union of items seen along all previous data blocks $D_1, \dots, D_t$.
%\hs{could $\mathcal{I}_{\text{neg}}$ include items appeared in previous user sequences?}, 

% \par\vspace{-0.04cm}\noindent
\smallsection{Inference.}
We compute relevance scores $ \sigma (\mathbf{E}_{\mathcal{I}^{1:t}}\mathbf{h}^L_N) \in [0,1]^{|\mathcal{I}^{1:t}|}$ by taking the dot product between the item embeddings $\mathbf{E}_{\mathcal{I}^{1:t}} \in \mathbb{R}^{|\mathcal{I}^{1:t}| \times d} $ and the hidden state at the last position \(\mathbf{h}^L_N \in \mathbb{R}^d\), which encodes information across all positions~\cite{bert4rec}.
These scores are then used to rank items for next item prediction.

% evaluate recommendation performance via various metrics (Hit, MRR, and NDCG).

% During training, the model predicts the next item at each position in the sequence, with the final prediction computed as $\mathbf{\hat{R}} = \mathbf{H}^L \mathbf{E}_\mathcal{I}^\top \in \mathbb{R}^{N \times |\mathcal{I}|}$, where $\mathbf{E}_\mathcal{I} \in \mathbb{R}^{|\mathcal{I}| \times d}$ is the item embedding matrix. In the inference stage, only the last hidden state $\mathbf{H}^L[-1]$ is used to predict the next item following the sequence.
% Despite its effectiveness, the quadratic time complexity for computing self-attention results in substantial computational costs as the sequence length increases.

% \vspace{-0.1cm}
\subsubsection{\textbf{Linear Attention.}}
% 핵심: 과거 데이터에 대한 접근 없이도 long-term depedencies를 caputre할 수 있다. 어떻게? self-attention에 대한 approximation을 수행하는데, 과거부터 누적되어온 hidden states와 현재 postion에서의 값과의 적절한 연산을 수행.. 여기서는 그게 도대체 뭔지를 보여주려고 한다.
Linear attention~\cite{linear_attention, linformer, performer} was originally designed to approximate self-attention in linear time—reducing its quadratic $\mathcal{O}(N^2)$ to $\mathcal{O}(N)$.
More recently, it has been extended to store past knowledge in parametric memories, thereby mitigating forgetting~\cite{Infini-attention, lee2023recasting}.
Given the ($l-1$)-th layer hidden states $\mathbf{H}^{l-1} \in \mathbb{R}^{N \times d}$, we form $\mathbf{Q} = \mathbf{H}^{l-1}\mathbf{W}_Q, \mathbf{K} = \mathbf{H}^{l-1}\mathbf{W}_K, \mathbf{V} = \mathbf{H}^{l-1}\mathbf{W}_V \in \mathbb{R}^{N \times d}$ (omitting the head‑specific index for clarity) and denote their $i$-th rows by $\mathbf{q}_i, \mathbf{k}_i, \mathbf{v}_i \in \mathbb{R}^d$, respectively.
A standard self-attention head computes $[\mathbf{a}_1, \dots, \mathbf{a}_N]^\top =\text{softmax}\left(\frac{\mathbf{Q}\mathbf{K}^\top}{\sqrt{d}}\right) \mathbf{V} \in \mathbb{R}^{N \times d}$.
In linear attention, the output at the $i$-th position $\mathbf{a}_i \in \mathbb{R}^d$ is computed:
% \vspace{-\topsep}
\begin{equation}
    \begin{aligned}
    \mathbf{a}_i =  \frac{\sum_{j=1}^{i}\text{sim}(\mathbf{q}_i, \mathbf{k}_j)\mathbf{v}_j}{\sum_{j=1}^{i} \text{sim}(\mathbf{q}_i, \mathbf{k}_j)}
    \end{aligned}\vspace{-\topsep}
    \label{eq:QKV_att}
\end{equation}
$\text{sim}(\mathbf{q}_i, \mathbf{k}_j)$ is expressed as $\phi(\mathbf{q}_i)^\top\phi(\mathbf{k}_j)$, using a kernel feature map function $\phi$ (e.g., ELU~\cite{elu}). 
Subsequently, Eq.\eqref{eq:QKV_att} is rewritten as:
%\vspace{-\topsep}
\small
\begin{equation}
\hspace{-0.4cm}
    \mathbf{a}_i =\frac{\sum^i_{j=1}\phi(\mathbf{q}_i)^\top\phi(\mathbf{k}_j)\mathbf{v}_j}{\sum^i_{j=1}\phi(\mathbf{q}_i)^\top\phi(\mathbf{k}_j)}
    = \frac{\phi(\mathbf{q}_i)^\top \sum^i_{j=1}\phi(\mathbf{k}_j)\mathbf{v}^\top_j}{\phi(\mathbf{q}_i)^\top \sum^i_{j=1}\phi(\mathbf{k}_j)} = \frac{\phi(\mathbf{q}_i)^\top \mathbf{s}_i}{\phi(\mathbf{q}_i)^\top 
    \mathbf{z}_i},
    \label{eq:linear_att}% \vspace{-0.2cm}
\end{equation}
\normalsize
\begin{equation}
\text{where} \quad
    \begin{aligned}
        \mathbf{s}_i = \sum_{j=1}^{i} \phi(\mathbf{k}_j) \mathbf{v}_j^\top, \quad \mathbf{z}_i = \sum_{j=1}^{i} \phi(\mathbf{k}_j)%.
    \end{aligned}% \vspace{-0.1cm}
    \label{eq:sz}
\end{equation}
Here, $\mathbf{s}_i \!\in\! \mathbb{R}^{d \times d}$ is called attention memory that encapsulates keys and values. 
Meanwhile, $\mathbf{z}_i \!\in\! \mathbb{R}^{d}$ is called normalizer memory that contains only keys.
% Meanwhile, $\mathbf{z}_i =\sum_{j=1}^{i} \phi(\mathbf{k}_j)\in \mathbb{R}^{d}$ is called as the normalizer memory, which includes only keys.
These memories are sequentially updated by combining the memories from the previous position (i.e., $\mathbf{s}_{i-1} \text{ and } \mathbf{z}_{i-1}$) with the current values (i.e., $\phi(\mathbf{k}_i) \mathbf{v}_i^\top \text{ and } \phi(\mathbf{k}_i)$), similar to hidden state propagation in RNNs:
\begin{equation}
    \begin{aligned}
        \mathbf{s}_{i} &= \mathbf{s}_{i-1} + \phi(\mathbf{k}_i)\mathbf{v}^\top_i,
        \quad
        \mathbf{z}_{i} = \mathbf{z}_{i-1} + \phi(\mathbf{k}_i)
    \end{aligned}
    \label{eq:sz_rnn}
\end{equation}
Note that $\mathbf{s}_{i} \text{ and } \mathbf{z}_{i}$ accumulate knowledge from position $1$ to $i$ in the input sequence $S^t_u$, serving as parametric memories~\cite{Infini-attention}.
Through these memories, linear attention partially retains historical knowledge without direct access to all previous positions~\cite{lee2023recasting}, while approximating self‑attention’s expressiveness.%\hs{Is this about previous "time stage" or just within the user sequences at time t?}

\vspace{-0.1cm}
\section{CSTRec}
\label{sec:method}

\begin{figure*}[t]
    \centering    
    \includegraphics[scale=1.4]{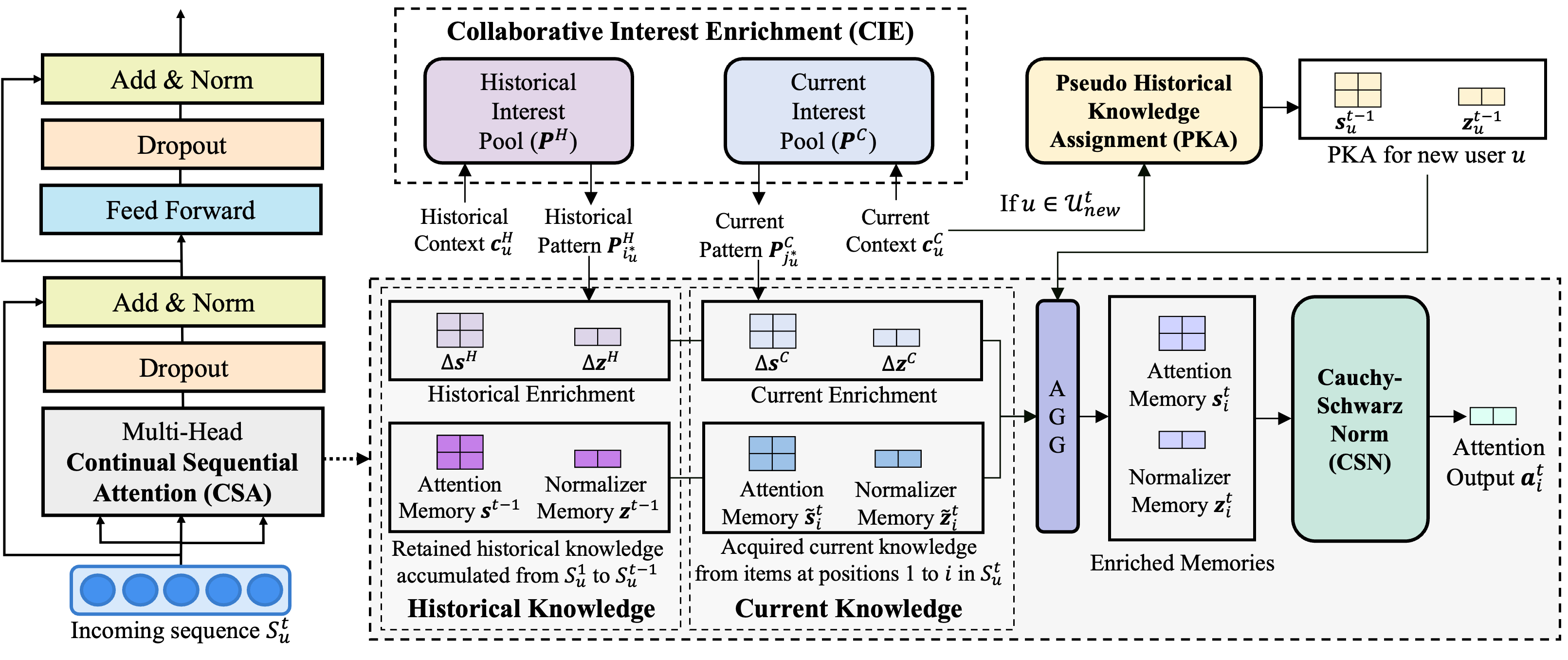}
    \caption{Overview of \proposed,
    illustrating the computation of attention output $\mathbf{a}^t_i$ for the $i$-th item in the incoming sequence $S^t_u$.
    Following the Multi-head CSA, we apply 
    the same transformer sublayer architecture (\cref{subsub:transformer})—dropout, layer normalization, and the position-wise feed-forward network. 
    AGG denotes the aggregation of historical and current knowledge in Eq.~(\ref{eq:provide_enrich}).
    }
    \label{fig:method}
    \vspace{-0.4cm}
\end{figure*}
We present \proposed, designed to effectively update Transformer-based SR model with data streams.
We begin by discussing how linear attention can be applied naively to continual SR (\cref{sub:overview}).
Next, we present Continual Sequential Attention, a specialized linear attention for continual SR (\cref{sub:CSA}), and introduce a new technique to facilitate accommodation of new users (\cref{sub:pseudo_h_new_users}).
Lastly, we present the optimization process (\cref{sub:objective_function}).
Figure~\ref{fig:method} shows an overview of \proposed, with main notations summarized in Table~\ref{tab:notation}.

% The proces for CSTRec is outlined in 
% Algorithm~\ref{al1} (Appendix~\ref{APP:algo}).

% We propose a novel \textbf{\underline{C}}ontinual \textbf{\underline{S}}equential \textbf{\underline{R}}ecommendation (\textbf{\proposed}) framework that extends CL to lifelong SR in a current environment. To effectively capture the user's evolving sequential patterns over time, CSR primarily comprises \textbf{continual sequential attention (\cref{sub:CSA})} and \textbf{historical and current interest pools (\cref{sub:pool})}. Additionally, we introduce a method to model \textbf{pseudo-historical knowledge for new users (\cref{sub:pseudo_h_new_users})}. An overview of \proposed is presented in Figure \ref{fig:method}.

\vspace{-0.1cm}
\subsection{Applying~Linear~Attention~to~Continual~SR}
\label{sub:overview}
% 목적.. input/output 및 보유하고 있는 과거지식 등 전반적인 상황을 설명해주자, training / inference 등?
% 이 section에서 우리는 \proposed가 과거지식과 현재지식을 어떻게 다뤄나가는지 보여주고자 간략하게 먼저 
We describe the operation of linear attention in continual SR through five steps: 
(1) retaining historical knowledge, (2) acquiring current knowledge, (3) integrating both knowledge, (4) computing attention, and (5) updating historical knowledge.
Assume that the model has been trained up to data block $D_{t-1}$, and is now being updated with data block $D_t$.
We omit the user index $u$ for notational simplicity.
\begin{enumerate}[leftmargin=*]
    \item \textbf{Retaining historical knowledge.}
    %step1
    At this point, the model stores historical knowledge in two parametric memories per user: attention memory $\mathbf{s}^{t-1}$ and normalizer memory $\mathbf{z}^{t-1}$. 
    The superscript denotes the time axis, indicating knowledge up to $D_{t-1}$.   
    % At this point, the model stores historical knowledge in two parametric memories per user: attention memory $\mathbf{s}^{t-1}$ and normalizer memory $\mathbf{z}^{t-1}$, which encompass knowledge up to $D_{t-1}$.
    Note that these memories are continuously updated along the data stream (i.e., $\mathbf{s}^1 \rightarrow \cdots \rightarrow \mathbf{s}^{t-2} \rightarrow \mathbf{s}^{t-1}$).
    \item \textbf{Acquiring current knowledge.}
    When an incoming user sequence $S^t_u$ arrives, 
    the model encodes current knowledge into attention memory $\Tilde{\mathbf{s}}^t_i$ and normalizer memory $\Tilde{\mathbf{z}}^t_i$ as follows:
    %수식1
    \vspace{-\topsep}
    \begin{equation}\label{eq:acq}
        \begin{aligned}
            \Tilde{\mathbf{s}}^t_i = \sum_{j = 1}^{i} \phi(\mathbf{k}^t_j)(\mathbf{v}^t_j)^\top, \quad
            \Tilde{\mathbf{z}}^t_i = \sum_{j = 1}^{i} \phi(\mathbf {k}^t_j)
        \end{aligned}\vspace{-\topsep}
    \end{equation}

    Note that these current memories (i.e., $\Tilde{\mathbf{s}}^t_i \text{ and } \Tilde{\mathbf{z}}^t_i$) capture only up-to-date knowledge from the first to the $i$-th item in $S^t_u$.
    \item \textbf{Integrating historical and current knowledge.} 
    We combine the attention and normalizer memories from both historical and current knowledge into $\mathbf{s}^{t}_i$ and $\mathbf{z}^{t}_i$, respectively, as follows:
    %step3
    \vspace{-0.1cm}
    \begin{equation}
        \begin{aligned}
            \mathbf{s}^{t}_i &= \mathbf{s}^{t-1} + \Tilde{\mathbf{s}}^t_i,
            \quad
            \mathbf{z}^{t}_i = \mathbf{z}^{t-1} + \Tilde{\mathbf{z}}^t_i
        \end{aligned}
        \label{eq:integrate}
    \end{equation}
    % Now, $\mathbf{s}^{t}_i \text{ and } \mathbf{z}^{t}_i$ reflect both historical and current knowledge.
    %step4
    \item \textbf{Computing attention output.}
    Finally, the attention output at the $i$-th position $\mathbf{a}^t_i \in \mathbb{R}^{d}$ is computed by rewriting Eq.~\eqref{eq:linear_att}:
    \vspace{-0.1cm}
    \begin{equation}
        \begin{aligned}
            \mathbf{a}^t_i &= \frac{\phi(\mathbf{q}^t_i)^\top \mathbf{s}^t_i}{\phi(\mathbf{q}^t_i)^\top \mathbf{z}^t_i}
        \end{aligned}
        \label{eq:att_out}
    \end{equation}
    This attention output reflects the comprehensive context of both historical and current knowledge, allowing \proposed to capture long-term user preferences along data streams.

    % step5:
    \item \textbf{Updating historical knowledge.} After training on $D_t$, the model updates two parametric memories per user as follows: 
    \vspace{-0.1cm}
    \begin{equation}
        \begin{aligned}
            \mathbf{s}^{t} &= \mathbf{s}^{t-1} + \Tilde{\mathbf{s}}^{t}_{N},
            \quad
            \mathbf{z}^{t} = \mathbf{z}^{t-1} + \Tilde{\mathbf{z}}^{t}_{N}, \\
            %&\text{where } \mathbf{s}^{1} = \mathbf{s}^{1}_{|S^{1}_u|} \text{ and } \mathbf{z}^{1} = \mathbf{z}^{1}_{|S^{1}_u|}.
        \end{aligned}\vspace{-0.1cm}
        \label{eq:update_his}
    \end{equation}
    where $\Tilde{\mathbf{s}}^{t}_{N} = \sum_{j = 1}^{N} \phi(\mathbf{k}^t_j)(\mathbf{v}^t_j)^\top \text{ and } \Tilde{\mathbf{z}}^{t}_{N} = \sum_{j = 1}^{N} \phi(\mathbf {k}^t_j)$ are the memories at the last position of the $N$-item sequence \(S^t_u\).
    % Note that 
    The updated memories $\mathbf{s}^t \text{ and } \mathbf{z}^t$ remain frozen during training.
\end{enumerate}
However, this naive application yields suboptimal results due to two main challenges:
First, \textbf{unstable learning} arises from an imbalance in the number of interactions per user along data streams.
Because the memories accumulate item representations from the given sequence over time (Eq.~\eqref{eq:update_his}), their magnitudes grow increasingly large for active users while remaining relatively small for less active users.
This disproportion across users leads to wide variation in attention output magnitudes, with active users empirically producing larger values. 
This in turn makes optimization highly unstable (without CSN in Figure 2), and the instability worsens as more data accumulates over time.
Second, \textbf{inevitable forgetting} occurs as new data continuously update the memories, gradually causing them to forget previously acquired knowledge.
In the following sections, we introduce our solutions to these challenges.

\subsection{Continual Sequential Attention (CSA)}
\label{sub:CSA}
We propose CSA, a tailored linear attention mechanism for continual SR.
It comprises two key components:
(1) Cauchy-Schwarz Normalization to prevent unstable learning, and (2) Collaborative Interest Enrichment to alleviate the inevitable forgetting.

\subsubsection{\textbf{Cauchy-Schwarz Normalization (CSN)}}
\label{subsub:csn}
We introduce CSN, a simple yet effective normalization technique based on the Cauchy-Schwarz inequality.
CSN is applied during the computation of attention outputs to ensure stable learning.
Let $\mathbf{q}^\prime_i = \phi(\mathbf{q}^t_i)$ and $\mathbf{a}_{\text{orig}}=\mathbf{a}^t_i$. Then, Eq.~\eqref{eq:att_out} becomes $\mathbf{a}_{\text{orig}} = \frac{(\mathbf{q}^\prime_i)^\top \mathbf{s}^t_i}{(\mathbf{q}^\prime_i)^\top \mathbf{z}^t_i}$.
% 코시슈바르츠.
%\hs{If there is enough space, it would be clearer to explicitly state $\mathbf{a}^t_i = \frac{(\mathbf{q}^\prime_i)^\top \mathbf{s}^t_i}{(\mathbf{q}^\prime_i)^\top \mathbf{z}^t_i}$ as equation, or just in-line.}
%To normalize both accumulated memories (i.e., $\mathbf{s}^t_i \text{ and } \mathbf{z}^t_i$), we apply the Cauchy–Schwarz inequality to the denominator as:
Here, we apply the Cauchy–Schwarz inequality to impose an upper bound on the denominator, thereby scaling the attention output magnitudes:
% \vspace{-\topsep}
%\small
\begin{equation}
\begin{aligned}
%\hspace{-0.3cm} % 기본 들여쓰기 제거
    (\mathbf{q}^\prime_i)^\top \mathbf{z}^t_i 
    %= \sum_{m=1}^{d} \mathbf{q}^\prime_{im} \mathbf{z}^t_{im} 
    %\leq \sqrt{\sum_{m=1}^{d} (\mathbf{q}^\prime_{im})^2} \sqrt{\sum_{m=1}^{d} (\mathbf{z}^t_{im})^2}
    \leq \|\mathbf{q}^\prime_i\|_2 \|\mathbf{z}^t_i\|_2
\end{aligned}
% \vspace{-0.2cm}
\end{equation}
\normalsize
% Equality holds when $\frac{\mathbf{q}^\prime_{i1}}{\mathbf{z}^t_{i1}} = \cdots = \frac{\mathbf{q}^\prime_{id}}{\mathbf{z}^t_{id}}$.
% To satisfy this condition, we introduce a loss that minimizes the standard deviation of the~ratios:
% %\vspace{-0.5cm}
% \begin{equation}
%     \begin{aligned}
%     \mathcal{L}_{\text{std}} = \log \Bigg( \sum_{S^t_u \in D_t} \sum_{i=1}^{|S^t_u|} \text{StdDev}\Bigg(\Big\{\frac{\mathbf{q}'_{im}}{\mathbf{z}^t_{im}} \;\Big|\; m = 1, \dots, d \Big\}\Bigg) \Bigg).
%     \end{aligned}\vspace{-\topsep}
% \end{equation}
%With the above derivation, we rewrite $\mathbf{a}^t_i$ as follows:
By replacing the denominator in $\mathbf{a}_\text{orig}$ with its Cauchy–Schwarz upper bound, we obtain $\mathbf{a}_\text{csn}$ as follows: 
%\vspace{-\topsep}
\begin{equation}
%\mathbf{a}^t_i = \frac{(\mathbf{q}^\prime_i)^\top \mathbf{s}^t_i}{(\mathbf{q}^\prime_i)^\top \mathbf{z}^t_i} 
%\mathbf{a}_\text{orig} \geq 
\mathbf{a}_\text{csn} =
\left(\frac{\mathbf{q}^\prime_i}{\|\mathbf{q}^\prime_i\|_2}\right)^\top \left(\frac{\mathbf{s}^t_i}{\|\mathbf{z}^t_i\|_2}\right) = \hat{\mathbf{q}}_i^\top \mathbf{r}^t_i,
%\vspace{-\topsep}
\label{eq:csn_attention}
\end{equation}
where $\hat{\mathbf{q}}_i \in \mathbb{R}^{d}$ is the L2-normalized query, and $\mathbf{r}^t_i \in \mathbb{R}^{d \times d}$ is CSN attention memory, respectively.
In terms of L2 norms, 
$\|\mathbf{a}_{\text{csn}}\|_2 = |\text{cos}{\theta_{qz}}|\cdot\|\mathbf{a}_{\text{orig}}\|_2, ~\text{where } 
    \text{cos}\theta_{qz}
    =\frac{(\mathbf{q}^\prime_i)^\top \mathbf{z}^t_i}{\|\mathbf{q}^\prime_i\|_2 \|\mathbf{z}^t_i\|_2} 
    =\frac{\mathbf{a}_{\text{csn}}}{\mathbf{a}_{\text{orig}}} 
$.
% Since $0 \leq |\text{cos}\theta_{qz}| \leq 1$, 
% $\mathbf{a}_{\text{csn}}$ prevents excessive divergence of $\mathbf{a}_{\text{orig}}$, keeping them within a stable range while maintaining the original relative magnitudes.
% Figure~2 shows that CSN effectively scales the magnitudes of attention outputs, enabling stable and effective optimization while taking advantage of linear attention. 
As $0 \leq |\text{cos}\theta_{qz}| \leq 1$, $\mathbf{a}_{\text{csn}}$ adaptively adjusts the magnitude between 0 and the original scale, preventing excessive divergence and keeping them within a stable range.
As $\mathbf{a}_{\text{csn}}$ only modifies the overall magnitude, it maintains the information in $\mathbf{a}_{\text{orig}}$.
Figure~2 shows that CSN effectively scales the magnitudes, enabling stable and effective optimization while taking advantage of linear attention. 
\begin{figure}[t!]
  \includegraphics[width=0.485\columnwidth]{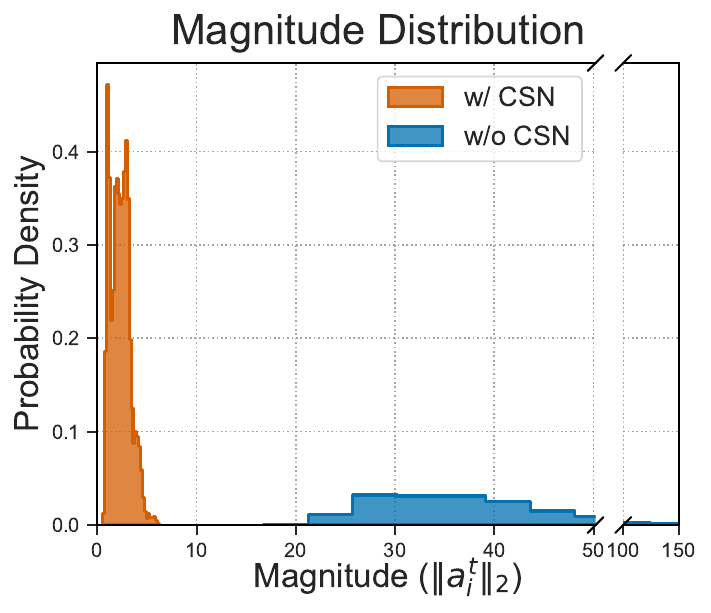}
  \includegraphics[width=0.505\columnwidth,
  height=0.16\textheight]{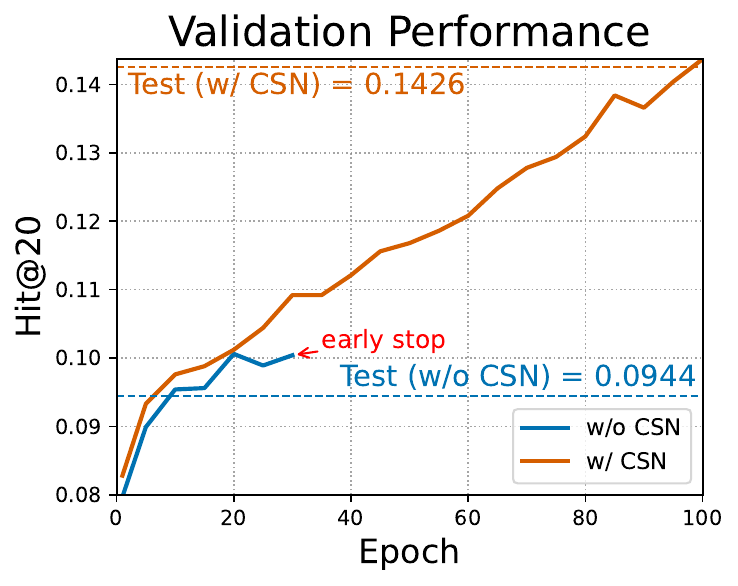}
  \caption{Effects of CSN on Yelp. 
  (Left) Magnitude distribution of linear attention output. 
  (Right) Validation performance.
  For `w/o CSN', we apply layer normalization to the attention outputs to promote stable training.}
  % with CSN, linear attention steadily improves; without CSN (), it quickly plateaus.}   
  % \caption{Effects of CSN on Yelp. 
  % (Left) Magnitude distribution of linear attention output. % on $D_4$.
  % (Right) Validation performance % on $D_2$
  % : with CSN, linear attention  steadily improves; without CSN (we apply layer normalization to attention outputs instead), it quickly plateaus.} 
  \label{fig:_interest_pool}
\end{figure}

% \vspace{-0.2cm}
\subsubsection{\textbf{Collaborative Interest Enrichment (CIE)}}
\label{sub:CIE}
We propose CIE to alleviate the inevitable forgetting and enhance the learning process.
% caused by continuous updates in Eq.~\eqref{eq:update}.
As its core, CIE introduces two types of \textit{interest pools}, each comprising a small set of learnable key–pattern pairs (i.e., $\mathcal{P} = \{({\kappa}_i, \mathbf{P}_i)\}^M_{i=1}$), where each key $\kappa_i \in \mathbb{R}^d$ is used for matching, and each pattern $\mathbf{P}_i \in \mathbb{R}^{L \times d}$ encodes a distinct aspect of user interests.
By leveraging this diverse interest knowledge, these pools are utilized to enrich the memories (i.e., $\mathbf{s}^t_i$ and $\mathbf{z}^t_i$) by selecting the most relevant interest for each user context.

% The motivations for introducing interest pools are twofold.
% First, the pools serve as external parametric memories that store key aspects of past knowledge and are updated more selectively than the main model parameters, thereby mitigating forgetting.
% Second, the pools are globally shared across users, enabling those with similar interests to access and leverage overlapping information.
% This generates collaborative signals that enhance the learning process.
% % When a pattern is repeatedly accessed by users with similar interests, the resulting updates are more consistent, which stabilizes its representation and helps reduce forgetting.
% Repeated access by users with similar interests yields consistent updates, stabilizing the representation and mitigating forgetting.

The motivations for introducing interest pools are twofold.
First, unlike the backbone transformer parameters that are frequently updated through next-item prediction, each pool updates only the most relevant pattern for a given user context. 
This selective update preserves specific aspects of past knowledge and reduces overwriting.
Second, the pools are globally shared across users, allowing similar users to collectively reinforce these patterns. 
As a result, they become more robust to forgetting and allow for leveraging recent behaviors of similar users when recalling historical patterns.

% CIE leverages two types of interest pools, focusing on (1) historical interests to compensate for forgetting, and (2) current interests to enhance the learning of new knowledge.
% Let $\mathbf{P}^H \text{ and } \mathbf{P}^C$ denote the historical and current interest pools. 

\smallsection{Historical and current interest pools.}
% \hs{would it be better to briefly mention what the key and value store when introducing the interest pools (maybe in 1-2 lines)? Otherwise, readers have to keep several new symbols in mind with no clue for 10+ lines.}
CIE leverages two interest pools: (1) Historical interest pool ($\mathcal{P}^H$), focusing on past interests to complement historical knowledge. (2) Current interest pool ($\mathcal{P}^C$), focusing on new interests to enrich current knowledge.
Simply put, $\mathcal{P}^H$ aims to preserve long-term preferences, while $\mathcal{P}^C$ supplements sparse information on emerging user interests, thereby ensuring the model covers both historical and current knowledge.

We employ a key-pattern pair design \cite{L2P, van2017neural, dualprompt}, where relevant patterns are retrieved through a matching process.
Each pool holds:
\begin{equation}
    \begin{aligned}
        \mathcal{P}^H &= \{ (\mathbf{\kappa}^H_i, \mathbf{P}^H_i)\}^{N_H}_{i=1}, \quad \mathbf{\kappa}^H_i \in \mathbb{R}^d, \quad\mathbf{P}^H_i \in \mathbb{R}^{L_H \times d}, \\
        \mathcal{P}^C &= \{ (\mathbf{\kappa}^C_j, \mathbf{P}^C_j)\}^{N_C}_{j=1}, \quad \mathbf{\kappa}^C_j \in \mathbb{R}^d, \quad\mathbf{P}^C_j \in \mathbb{R}^{L_C \times d}
    \end{aligned}
\end{equation}
$N$ and $L$ are hyperparameters that determine each pool's capacity: $N$ specifies the number of interests stored (variety of knowledge), and $L$ specifies the length of each interest (level of detail). 
We provide a detailed study in \cref{subsub:hyper}.

%For each pool, $N_H, N_C$ are the numbers of interests, while $L_H, L_C$ are the length of each interest.
% These hyperparameters decide the pool capacity. We provide a detailed study~in~\cref{subsub:hyper}.

%Each key $\mathbf{\kappa}^H_i,\mathbf{\kappa}^C_j$ serves as an identifier for the corresponding values $\mathbf{P}^H_i,\mathbf{P}^C_j$ and is used to retrieve relevant values given contexts $\mathbf{c}^H_u, \mathbf{c}^C_u \in \mathbb{R}^d$ , respectively.
% pool parameters
% \footnote{These hyperparameters decide the pool capacity. We provide a detailed study~in~\cref{subsub:hyper}.}

\smallsection{Interest enrichment.}
Given an input sequence $S^t_u = [i^t_1, \dots, i^t_N]$, we obtain historical and current contexts $\mathbf{c}^H_u, \mathbf{c}^C_u \in \mathbb{R}^d$ from its last hidden state from the final Transformer layer.\footnote{The computation follows the Transformer layer with CSA (\cref{subsub:transformer}). 
During the attention computation in Eq. \eqref{eq:csn_attention}, we use ($\Tilde{\mathbf{s}}^t_i, \Tilde{\mathbf{z}}^t_i$) for $\mathbf{c}^C_u$ to better focus on the current interest, while leveraging (${\mathbf{s}}^t_i, {\mathbf{z}}^t_i$) to reflect both historical and current aspects for $\mathbf{c}^H_u$.}
%\hs{ This $\mathbf{q}$ is confusing with the $\mathbf{q}$ in Eq.4. Also, is this the same as $\mathbf{h}^L_N$ in Eq.2?}
Using these contexts, we retrieve the most relevant interest in each pool. 
%Here, we explain the process of historical interests, and the same process applies to current interests.
For $\mathbf{c}^H_u$ and $\mathbf{c}^C_u$, we identify the best‐matching indices $i^*_u$ and $j^*_u$ in pools $\mathcal{P}^H$ and $\mathcal{P}^C$, respectively, using a matching function $\gamma$.\footnote{In this work, we use cosine distance for its simplicity.}
\begin{equation}
    \begin{aligned}
        i^*_u &= \underset{i\in\{1,\dots,N_H\}}{\text{argmin}} \gamma(\mathbf{c}^H_u, \mathbf{\kappa}^H_i), \quad j^*_u = \underset{j\in\{1,\dots,N_C\}}{\text{argmin}} \gamma(\mathbf{c}^C_u, \mathbf{\kappa}^C_j)
    \label{eq:query}
    \end{aligned}
\end{equation}
%Similarly, for $\mathbf{c}^C_u$ we find $j^*_u$ in the current pool $\mathcal{P}^C$.
Note that we store the best-matching indices and use them at the inference phase, enabling interest enrichment with~negligible~costs.
% historical and current interest enrichment
The corresponding interest knowledge $\mathbf{P}^H_{i^*_u} \text{ and } \mathbf{P}^C_{j^*_u}$ form historical interest enrichment $(\Delta\mathbf{s}^H, \Delta\mathbf{z}^H)$ and current interest enrichment $(\Delta\mathbf{s}^C, \Delta\mathbf{z}^C)$, respectively.
These are computed using the key- and value-projection matrices $\mathbf{W}_K$ and $\mathbf{W}_V$, as in Eq.~\eqref{eq:sz}:
%\hs{maybe good idea to remind/recall what $W_K, W_V$ are?}
%\hs{btw, so what is $\mathbf{k}$ and $\mathbf{P}$..?}
\vspace{-\topsep}
%\small
\begin{equation}
%\hspace{-0.6cm}
    \begin{aligned}
    \Delta\mathbf{s}^H &= \sum_{l = 1}^{L_H} \phi( \{\mathbf{P}^H_{i^{*}_{u}}\![l,:]\,\mathbf{W}_K)(\mathbf{P}^H_{i^{*}_{u}}\![l,:]\,\mathbf{W}_V )^\top, \;
    \Delta\mathbf{z}^H = \sum_{l = 1}^{L_H} \phi(\mathbf{P}^H_{i^{*}_{u}}\![l,:])\mathbf{W}_K \\
    \Delta\mathbf{s}^C &= \sum_{l = 1}^{L_C} \phi( \mathbf{P}^C_{j^{*}_{u}}\![l,:]\,\mathbf{W}_K)(\mathbf{P}^C_{j^{*}_{u}}\![l,:]\,\mathbf{W}_V )^\top, \;
    \Delta\mathbf{z}^C = \sum_{l = 1}^{L_C} \phi(\mathbf{P}^C_{j^{*}_{u}}\![l,:])\mathbf{W}_K
    \end{aligned}\vspace{-\topsep}
    \label{eq:get_enrich}
\end{equation}
%\normalsize
We then extend Eq.~\eqref{eq:integrate} to aggregate both aspects as follows: 
% \vspace{-\topsep}
\begin{equation}
\begin{aligned}
\mathbf{s}^t_i 
  &= \underbrace{\mathbf{s}^{t-1} + \Delta\mathbf{s}^H}_{\text{Historical knowledge}}
     \;+\;
     \underbrace{\tilde{\mathbf{s}}^t_i + \Delta\mathbf{s}^C}_{\text{Current knowledge}}\\
\mathbf{z}^t_i 
  &= \underbrace{\mathbf{z}^{t-1} + \Delta\mathbf{z}^H}_{\text{Historical knowledge}}
     \;+\;
     \underbrace{\tilde{\mathbf{z}}^t_i + \Delta\mathbf{z}^C}_{\text{Current knowledge}}
    \label{eq:provide_enrich}
\end{aligned}
\end{equation}
The enriched memories $\mathbf{s}^t_i \text{ and } \mathbf{z}^t_i $ from Eq.~\eqref{eq:provide_enrich} are then passed to the CSN step in Eq.~\eqref{eq:csn_attention}.
Lastly, we introduce the following loss for accurate matching: 
\begin{equation}
    \begin{aligned}
        \mathcal{L}_{\text{match}}= \frac{1}{|D_t|}\sum_{S^t_u \in D_t} \left[ \gamma(\mathbf{c}^H_u, \mathbf{\kappa}^H_{i^*_u}) + \gamma(\mathbf{c}^C_u, \mathbf{\kappa}^C_{j^*_u}) \right],
    \end{aligned}
    \label{eq:match}
\end{equation}
%\hs{why is this loss averaged over dataset but not bce (Eq.2)?}
% This pull selected keys closer to corresponding queries, allowing for progressively capturing diverse interests.
%where $i^*_u, j^*_u$ denote the best matching indices for each historical and current query in Eq.~\eqref{eq:query}.
This loss function pulls selected keys closer to corresponding contexts, allowing for progressively capturing more accurate interests.
As a result, historical and current interests encode distinguishable knowledge.
Please refer to \cref{subsub:interest_pool_analysis} for further analysis.

\subsection{Pseudo-Historical Knowledge Assignment}
\label{sub:pseudo_h_new_users}
One critical challenge in continual SR is accommodating newly joined users who have no past interactions. %, known as the user cold-start problem.
At time period $t$, new users have current memories ($\tilde{\mathbf{s}}^t_i, \tilde{\mathbf{z}}^t_i$), but has no historical ones ($\mathbf{s}^{t-1}, \mathbf{z}^{t-1}$).
To address this, we introduce a pseudo-historical knowledge assignment that leverages the historical knowledge of existing users with the most similar behavioral patterns.
For each new user, the process involves two steps: (1) identifying the top-$K$ existing users with similar current interests, and (2) assigning pseudo-historical knowledge derived from their historical knowledge.

First, for each new user $u$, we identify top-$K$ existing users $\mathcal{N}_u$:
\begin{equation}\label{eq:topk}
    \begin{aligned}
            \mathcal{N}_u &=\{u^\prime | \underset{u^\prime \in \mathcal{U}^{t}\setminus \mathcal{U}^{t}_{\text{new}}}{\operatorname{argsort}}  \gamma(\mathbf{c}^C_u, \mathbf{c}^C_{u^\prime})[: K]\},
    \end{aligned}
\end{equation}
where $\mathcal{U}^{t}\setminus \mathcal{U}^{t}_{\text{new}}$ is the set of existing users in $D_{t}$, excluding new users. 
Here, $\mathbf{c}^C_u$ is the context of current interest, obtained from the last hidden state of $S^t_u$ (\cref{sub:CIE}).
Next, we generate pseudo-historical knowledge, weighted by their similarity scores $\psi(u, u')$ as:
% \vspace{-\topsep}
\begin{equation}
        \mathbf{s}^{t-1}_u = \sum_{u' \in \mathcal{N}_u} \psi(u, u^\prime) \mathbf{s}^{t-1}_{u^\prime}, \quad
        \mathbf{z}^{t-1}_u = \sum_{u' \in \mathcal{N}_u} \psi(u, u^\prime) \mathbf{z}^{t-1}_{u^\prime},
    \label{eq:pseudo_history}% \vspace{-0.15cm}
\end{equation}
where $\mathbf{s}^{t-1}_u$ and $\mathbf{z}^{t-1}_u$ are the pseudo-historical knowledge for the new user $u$.
% We use the Softmax function to obtain normalized weight, i.e., $\psi(u, u^\prime) = \frac{\exp\left(\mathbf{c}^C_u \cdot \mathbf{c}^C_{u'} / \tau \right)}{\sum_{u' \in \mathcal{N}_u} \exp\left(\mathbf{c}^C_u \cdot \mathbf{c}^C_{u'} / \tau\right)}$.
We use the Softmax function to obtain normalized weight.\footnote{$\psi(u, u^\prime) = \exp(\mathbf{c}^C_u \cdot \mathbf{c}^C_{u'} / \tau )   /  \sum_{u' \in \mathcal{N}_u} \exp(\mathbf{c}^C_u \cdot \mathbf{c}^C_{u'} / \tau)$} 
Now, $\mathbf{s}^{t-1}_u$ and $\mathbf{z}^{t-1}_u$ serve as historical knowledge for CSA computation.
% This technique complements the lack of historical data for new users, facilitating their learning and adaptation in \proposed.
This technique complements the lack of historical data for new users, facilitating their adaptation in \proposed.

\begin{algorithm}[t]
\footnotesize
\DontPrintSemicolon
\SetKwInOut{Input}{Input}
\SetKwInOut{Output}{Output}

\Input{Data block $D_t = \{S_u^t\}_{u \in \mathcal{U}^t}$, Model $M(\cdot;\theta)$, 
Pools $\mathcal{P}^H, \mathcal{P}^C$, 
Memories $\{\mathbf{s}^{t-1}_u, \mathbf{z}^{t-1}_u \mid u \in \mathcal{U}^{t} \setminus \mathcal{U}^{t}_\text{new} \}$
}

\Output{Updated model $M(\cdot;\theta)$, Updated Pools $\mathcal{P}^H, \mathcal{P}^C$, Updated Memories $\{\mathbf{s}^{t}_u, \mathbf{z}^{t}_u \mid u \in \mathcal{U}^{t}\}$
}

\BlankLine

\SetInd{0.2em}{0.8em}
\For{\textup{each epoch}}{
    \SetInd{0.2em}{0.8em} % 들여쓰기 결정.. 첫번째 indent, 그 다음 indent

    \If{\textup{epoch} \% $R = 0$}{
        Assign pseudo-historical knowledge $\mathbf{s}^{t-1}_u, \mathbf{z}^{t-1}_u, \forall u \in\mathcal{U}^t_\text{new}$\Comment{Eq. \eqref{eq:pseudo_history}} \\
        Identify the best-matching
indices $i^*_u, j^*_u, \forall u \in \mathcal{U}^t$  \Comment{Eq. \eqref{eq:query}}
    }
% for each user
    \For{\textup{each \proposed layer}}{
        \SetInd{0.2em}{0.8em}

        \For{\textup{each head}}{
        Acquire current memories at the $i$-th position $\Tilde{\mathbf{s}}^{t}_i, \Tilde{\mathbf{z}}^{t}_i$ \Comment{Eq. \eqref{eq:acq}} \\
        Retrieve historical memories $\mathbf{s}^{t-1}, \mathbf{z}^{t-1}$ \\
        Derive $(\Delta\mathbf{s}^{H}, \Delta\mathbf{z}^{H}, \Delta{\mathbf{s}}^{C}, \Delta{\mathbf{z}}^{C})$ then enrich $(\mathbf{s}^t_i, \mathbf{z}^t_i)$ 
        \Comment{Eq.~\eqref{eq:get_enrich}, \eqref{eq:provide_enrich}} \\
        Compute attention output $\mathbf{a}^{t}_i$ using CSN  \hfill\Comment{Eq. \eqref{eq:csn_attention}} \\
        % Generate the final attention output $\mathbf{a}^t_i \in \mathbf{A}$ \hfill\Comment{Eq.\eqref{eq:a_hisnew}}
        % Aggregate both attention outputs into $\mathbf{a}^t_i \in \mathbf{A}$ \hfill\Comment{Eq.\eqref{eq:a_hisnew}}
        }
        % Approximate self-attention, followed by MH and FFN \Comment{Eq.\eqref{eq:self_att}}
        Aggregate multi-head results, followed by FFN, Dropout, LayerNorm % \Comment{Eq.\eqref{eq:self_att}}
    }
    Optimize the parameters by minimizing the loss\Comment{Eq.\eqref{eq:optimize}}
    % Compute recommendation scores: \\
    % $\mathbf{\hat{R}} = \mathbf{H}^L \mathbf{E}_\mathcal{I}^\top$, where $\mathbf{\hat{R}} \in \mathbb{R}^{N \times |\mathcal{I}|}$ and $\mathbf{E}_\mathcal{I} \in \mathbb{R}^{|\mathcal{I}| \times d}$ are item embeddings.
}
Update historical memories $\mathbf{s}^t_u, \mathbf{z}^t_u , \forall u \in \mathcal{U}^t$ 
\Comment{Eq.~\eqref{eq:update_his}}
%\BlankLine
\caption{\proposed algorithm on $t$-th data block ($D_t$)}
\label{al1}
\end{algorithm}

% \vspace{-0.05cm}
\subsection{Optimization of \proposed}
\label{sub:objective_function}
The final learning objective of \proposed is as follows:
% \vspace{-\topsep}
\begin{equation}
    \begin{aligned}
        \min_{\substack{\theta, \mathcal{P}^H, \mathcal{P}^C}} \mathcal{L}_\text{BCE} + \lambda_\text{match}\mathcal{L}_\text{match}
    \end{aligned}%\vspace{-0.15cm}
    \label{eq:optimize}
\end{equation}
%\hs{Is this BCE from Eq 2? Is it different from Eq 2? Would it be better to denote it differently or specify depending variables/parameters to emphasize that this loss is introduced by us}
% \par\vspace{-0.2cm} % 원하는 만큼 간격 설정
\noindent where $\theta, \mathcal{P}^H, \mathcal{P}^C$ are the parameters of \proposed and interest pools.
$\mathcal{L}_\text{BCE} \text{ and } \mathcal{L}_\text{match}$ are defined in Eqs.~\eqref{eq:bce} and ~\eqref{eq:match}, and $\lambda_\text{match}$ is a hyperparameter for balancing the losses.
The training process is detailed in Algorithm~\ref{al1}.
Pseudo knowledge assignment (line 3) and interest pool retrieval (line 4) are performed every $R$ epochs, as conducting them every epoch is unnecessary and time-consuming.

% $\lambda_\text{match}$ is a hyperparameter controlling the impact of the matching loss.
% At each training step, after assigning pseudo-historical knowledge to newly incoming users (\cref{sub:pseudo_h_new_users}) and identifying two types of learnable interests for each current user (\cref{sub:pool}), 

\smallsection{Time Complexity Analysis.}
For each transformer layer in \proposed, the CSA head with the CIE module requires $\mathcal{O}\bigl(Nd^2\bigr)$ for linear attention (Eq.~\eqref{eq:linear_att}) and $\mathcal{O}\bigl((L_H + L_C)d^2\bigr)$ for interest enrichment (Eq.~\eqref{eq:get_enrich}), yielding $\mathcal{O}\bigl((N+ L_H + L_C)d^2\bigr) \approx \mathcal{O}\bigl(Nd^2\bigr)$.
This further simplifies to $\mathcal{O}\bigl(N\bigr)$, since $N \!\gg\! d$ and $d$ is a fixed constant~\cite{linrec, limarec}.
Identifying the best-matching indices (Eq.~\eqref{eq:query}) and top-$K$ existing users (Eq.~\eqref{eq:topk}) requires $\mathcal{O}\bigl(|\mathcal{U}^t|(N_H + N_C)d\bigr)$ and $\mathcal{O}\bigl(|\mathcal{U}^t_\text{new}|(|\mathcal{U}^t| - |\mathcal{U}^t_\text{new}|)d\bigr)$, respectively. 
Both operations run only once every $R$ epochs—not per layer or per position—so their amortized overhead per epoch is negligible compared to the cost of a single transformer layer.

\vspace{0.001in}
\noindent
\textbf{Efficiency of \proposed.}
Compared to the self-attention, \proposed shows comparable efficiency for training and greatly reduced efficiency for inference via three key designs:
(1) Building upon linear attention, \proposed reduces computational complexity with respect to the input length from $\mathcal{O}(N^2)$ to $\mathcal{O}(N)$.
(2) During training, \proposed efficiently performs CIE and pseudo-historical knowledge assignment at predefined intervals $R$, reducing overhead.
(3) During inference, \proposed leverages the best-matching indices identified during the training, enabling efficient CIE with negligible costs.
A detailed analysis of efficiency is provided in Table~\ref{tab:att}.

\section{Experiments}
%\vspace{-0.1cm}
\subsection{Experimental Setup}
\label{sec:experimentsetup}
% We provide details of setup in Appendix \ref{App:setup}.

% a movie rating dataset, and  a location-based social networking dataset.

% \input{sections/main_table_FT_onetable}

% \input{sections/main_table_FB_onetable}

% \input{sections/main_table_H20}

% \vspace{-0.1cm}
%to reflect diverse characteristics in sequence lengths and interaction counts.
% For ML-1M and Yelp, we binarize ratings larger than $3$ as 1.
% For ML-1M and Yelp, we remove interactions with ratings less than 3.
% Note that each data block is composed of sequences of items. 
% \usepackage{multirow}

\begin{table}[t]
\caption{Data block statistics after preprocessing.}
\centering
\resizebox{\columnwidth}{!}{%
\begin{tabular}{cc|c|cccc}
\toprule
\multicolumn{2}{c|}{\textbf{Data Blocks}}                                                       & $\mathbf{D_0}\;(60\%)$           & $\mathbf{D_1}\;(10\%)$       & $\mathbf{D_2}\;(10\%)$         & $\mathbf{D_3}\;(10\%)$         & $\mathbf{D_4}\;(10\%)$         \\ \hline\hline
\multicolumn{1}{c|}{\multirow{5}{*}{\rotatebox[origin=c]{90}{\textbf{Gowalla}} }} & \textbf{\# of users (new users)} & 30,682(30,682) & 2,364(692) & 2,227(828)   & 2,334(902)   & 2,490(1,082) \\
\multicolumn{1}{c|}{}                                  & \textbf{\# of items (new items)} & 68,189(68,189) & 3,006(920) & 2,879(1,059) & 3,000(1,123) & 3,076(1,169) \\
\multicolumn{1}{c|}{}                                  & \textbf{\# of interactions}            & 1,754,145      & 49,637     & 45,738       & 46,956       & 49,127       \\
\multicolumn{1}{c|}{}                                  & \textbf{Avg. Seq Length}               & 57.17          & 21.00      & 20.54        & 20.12        & 19.73        \\
\multicolumn{1}{c|}{}                                  & \textbf{Sparsity}                      & 0.9992         & 0.9930     & 0.9929       & 0.9933       & 0.9936       \\ \hline
% ml-1m
\multicolumn{1}{c|}{\multirow{5}{*}{\rotatebox[origin=c]{90}{\textbf{ML-1M}} }}   & \textbf{\# of users (new users)} & 3,978(3,978)   & 820(777)   & 767(567)     & 844(597)     & 932(109)     \\
\multicolumn{1}{c|}{}                                  & \textbf{\# of items (new items)} & 2,845(2,845)   & 1,680(6)   & 1,768(4)     & 1,726(0)     & 1,876(18)    \\
\multicolumn{1}{c|}{}                                  & \textbf{\# of interactions}            & 498,877        & 77,470     & 77,382       & 76,933       & 76,619       \\
\multicolumn{1}{c|}{}                                  & \textbf{Avg. Seq Length}               & 125.41         & 94.48      & 100.89       & 91.15        & 82.21        \\
\multicolumn{1}{c|}{}                                  & \textbf{Sparsity}                      & 0.9559         & 0.9438     & 0.9429       & 0.9472       & 0.9562       \\ \hline
%yelp
\multicolumn{1}{c|}{\multirow{5}{*}{\rotatebox[origin=c]{90}{\textbf{Yelp}} }} & \textbf{\# of users (new users)} & 104,281(104,281) & 7,340(3,634) & 7,065(3,820) & 7,173(3,643) & 8,260(4,312) \\
\multicolumn{1}{c|}{}                                  & \textbf{\# of items (new items)} & 52,290(52,290) &  5,736(669) & 5,708(843)   & 6,283(1,093)   & 7,183(1,401) \\
\multicolumn{1}{c|}{}                                  & \textbf{\# of interactions}            & 1,449,055      & 66,365     & 64,364       & 68,152       & 84,820       \\
\multicolumn{1}{c|}{}                                  & \textbf{Avg. Seq Length}               & 13.89          & 9.04      & 9.11        & 9.50        & 10.26        \\
\multicolumn{1}{c|}{}                                  & \textbf{Sparsity}                      & 0.9997         & 0.9984     & 0.9984       & 0.9985       & 0.9986       \\ \bottomrule
\end{tabular}
}
\label{tab:datablocks}
\end{table}

% \begin{table}[t]
% \caption{Data block statistics of two datasets.}
% \centering
% \renewcommand{\tabcolsep}{1.1mm}
% \resizebox{\columnwidth}{!}{%
% \begin{tabular}{cc|c|ccccc}
% \hline
% \multicolumn{2}{c|}{\textbf{Data Blocks}} & $\mathbf{D_0}$ & $\mathbf{D_1}$ & $\mathbf{D_2}$ & $\mathbf{D_3}$ & $\mathbf{D_4}$ & $\mathbf{D_5}$ \\ \hline\hline
% \multicolumn{1}{c|}{} & \textbf{\# of accumulated users} & 19,668 & 21,266 & 24,107 & 26,840 & 29,106 & 29,858 \\
% \multicolumn{1}{c|}{} & \textbf{\# of accumulated items} & 39,354 & 39,809 & 40,381 & 40,691 & 40,908 & 40,988 \\
% \multicolumn{1}{c|}{} & \textbf{\# of new users} & -- & 1,598 & 2,841 & 2,733 & 2,266 & 752 \\
% \multicolumn{1}{c|}{} & \textbf{\# of new items} & -- & 455 & 572 & 310 & 217 & 80 \\
% \multicolumn{1}{c|}{\multirow{-5}{*}{\rotatebox[origin=c]{90}{\textbf{Gowalla}} }} & \textbf{\# of interactions} & 513,732 & 62,336 & 112,167 & 123,042 & 126,474 & 89,713 \\ \hline

% \multicolumn{1}{c|}{} & \textbf{\# of accumulated users} & 12,248 & 13,367 & 13,896 & 14,399 & 14,761 & 14,950 \\
% \multicolumn{1}{c|}{} & \textbf{\# of accumulated items} & 10,822 & 11,345 & 11,634 & 11,970 & 12,197 & 12,261 \\
% \multicolumn{1}{c|}{} & \textbf{\# of new users} & -- & 1,119 & 529 & 503 & 362 & 189 \\
% \multicolumn{1}{c|}{} & \textbf{\# of new items} & -- & 523 & 289 & 336 & 227 & 64 \\
% \multicolumn{1}{c|}{\multirow{-5}{*}{   \rotatebox[origin=c]{90}{\textbf{Yelp}}  }} & \textbf{\# of interactions} & 151,084 & 51,280 & 34,501 & 33,593 & 37,970 & 34,189 \\ \hline
% \end{tabular}
% }
% \label{tab:datablocks}
% \vspace{0.1cm}
% \end{table}

\subsubsection{\textbf{Datasets.}} 
We use three real-world datasets: Gowalla, ML-1M, and Yelp, following~\cite{lee2024continual, PIW, limarec, linrec}.
% Gowalla\footnote{\url{https://snap.stanford.edu/data/loc-gowalla.html}}, ML-1M\footnote{\url{https://grouplens.org/datasets/movielens/1m}}, and Yelp\footnote{\url{https://www.yelp.com/dataset}}.
To simulate non-stationary data streams, we split each dataset chronologically.
The first 60\% of the data serves as the base block ($D_0$), which is used to pretrain all methods before the continual updates begin.
The remaining 40\% is equally divided into four incremental blocks ($D_1$ to $D_4$), following prior CL studies \cite{GraphSAIL, mi2020ader, lee2024continual}.
We apply $k$-core filtering with $k = 5$ for Yelp and $k = 10$ for other datasets for each block.
Each interaction sequence in a block is split into training, validation, and test sets based on item positions: the last item serves as the test label, the second-to-last item as the validation label, and the remaining items are used for training.
Table \ref{tab:datablocks} presents detailed statistics.

%We evaluate the performance on  incremental blocks from $D_2$ to $D_4$.
% For the task of SR (i.e., next-item prediction), 

% \vspace{-0.1cm}
\subsubsection{\textbf{Evaluation metrics.}} 
Recent CL studies \cite{do2023continual, lee2024continual} have employed specialized metrics to evaluate a model's ability to retain and acquire knowledge over time.
Following these work, we adopt two CL metrics: \textit{Retained Average (RA)} and \textit{Learning Average (LA)}.
Specifically, we construct a performance matrix $A \in \mathbb{R}^{t \times t}$, where each entry $a_{i j}$ (with $i \geq j$) represents the recommendation performance on block $j$ after training on block $i$.
After updating the model on the $t$-th block (i.e., $D_t$), we report the following metrics:
\begin{itemize}[leftmargin=*]% \vspace{-\topsep}
    \item \textbf{RA} $: \frac{1}{t} \sum^{t}_{i=1} a_{t,i}$ evaluates knowledge retention from past blocks.
    \item \textbf{LA} $: \frac{1}{t} \sum^{t}_{i=1} a_{i,i}$ evaluates knowledge acquisition from new blocks.
\end{itemize}% \vspace{-\topsep}
\vspace{-0.2cm}
Additionally, we report \textit{H-mean}, the harmonic mean of LA and RA, to provide an overall comparison of a model's capability \cite{do2023continual}. 
We use Hit@20 (H@20), MRR@20 (M@20)~\cite{MRR}, and NDCG@20 (N@20)~\cite{NDCG} as evaluation metrics~\cite{GraphSAIL, PIW, do2023continual}.
All results are averaged over five independent runs with different random seeds.
Note that we report results after training on $D_2, D_3 \text{ and }D_4$ in Tables~\ref{tab:fine_tune} and~\ref{tab:full_batch}, as forgetting becomes evident at those stages, but not on $D_1$.

\vspace{-0.2cm}
\subsubsection{\textbf{Baselines}}
For a thorough evaluation, we adopt two distinct training setups and compare state-of-the-art methods for~each~setup.
%  the backbone for methods that do not propose a specific model architecture (e.g., SAIL-PIW~\cite{PIW} and Reloop2~\cite{zhu2023reloop2}).
% For thorough evaluation, we categorize the baselines into two CL setups: Fine-tune and Full-batch, described as follows: 
\begin{enumerate}[leftmargin=*]\vspace{-\topsep}
    \item \textbf{Fine-tune}: The model is continually updated using only the incoming data block, without direct access to historical data.
    We compare the following state-of-the-art CL methods:
    \begin{itemize}[leftmargin=*]
        \item \textbf{SAIL-PIW ~\cite{PIW}} is a regularization-based method that assigns regularization weights based on user preference shift.
        % based on whether user preferences is static or dynamic.
        \item \textbf{Reloop2~\cite{zhu2023reloop2}} is a replay-based method that uses an error memory module to improve future recommendations.
        \item \textbf{IMSR~\cite{IMSR}} is an incremental learning framework for SR, which utilizes multiple interest representations for each user.
        % is an incremental learning framework for multi-interest SR to accommodate users' new interests while preventing forgetting of their existing interests. 
        % is a framework for incremental multi-interest SR.
    \end{itemize}
    \item \textbf{Full-batch}: The model is updated with all incremental blocks. 
    % The model is updated with all historical data. 
    % For efficient training, we exclude the base block \(D_0\). 
    We compare recent methods designed to capture user interests formed over long periods (i.e., lifelong or long-term SR models):
    % We compare the following methods that effectively capture user preferences formed over long periods (i.e., lifelong or long-term SR models):
    \begin{itemize}[leftmargin=*]
        \item \textbf{HPMN~\cite{HPMN}} uses GRUs for personalized memorization to capture multi-scale sequential patterns in lifelong sequences.
        \item \textbf{LimaRec~\cite{limarec}} builds upon a linear attention mechanism to capture multi interests of users within lifelong sequences.
        % \item \textbf{LinRec~\cite{linrec}} proposes a normalized linear attention mechanism for long-term user behavior modeling.
        \item \textbf{LinRec~\cite{linrec}} proposes an L2-normalized linear attention mechanism that leverages dual-side normalization techniques.
        % for long-term user behavior modeling.
    \end{itemize}
\end{enumerate}\vspace{-\topsep}
As a representative transformer-based SR model, we use SASRec~\cite{sasrec} for performance comparison in both setups.
For baselines without their own architecture (i.e., SAIL-PIW and Reloop2), we integrate their modules into SASRec to evaluate their peformance on SR.

\vspace{-0.2cm}
\subsubsection{\textbf{Implementation details.}}
We utilize PyTorch with CUDA, utilizing RTX 3090 GPU and AMD EPYC 7413 CPU.
Hyperparameters are tuned through grid search on the validation set.
%lr
The learning rate is chosen from \{0.0001, 0.0002, 0.0005\}.
%L_2
$L_2$ regularization for Adam is chosen from \{5e-6, 1e-5, 5e-5, 1e-4\}, and
% dropout
the dropout ratio from
\{0.05, 0.075, 0.1, 0.125\}.
% head/layer/negatvie samples
The number of heads $h$, layers $L$, and negative samples are set to 2.
% dim, window_size
The dimension $d$ is set to $64$ for ML‑1M, $32$ for Gowalla, and $16$ for Yelp. 
In the fine‑tune setup, the window sizes are set to $50$, $25$, and $10$ for each dataset, respectively, and are doubled in the full‑batch setup to better capture long‑term dependencies. 
These choices align with the guideline of maintaining a ratio of sequence length to dimension greater than $1.5$ for long-term SR scenarios~\cite{linrec}, while also considering the average sequence lengths.
% The dimension is set to 32 for Gowalla, 64 for ML-1M, and 16 for Yelp.
% The window size for sequences is set to 25, 50, and 10, respectively.
% Note that we double the window size in full-batch setup to better capture long-term dependencies.
%considering that a ratio of length to dimension greater than 1.5 is regarded as long-term."
% interests
For \proposed, the number of interests $N_H, N_C$ is chosen from \{10, 20, ..., 50\}, interest lengths $L_H, L_C$ from \{10, 20, 50\}.
The number of similar users $K$ for pseudo-historical knowledge assignment is chosen from \{5, 10, 15, 20, 25\}. %and $\lambda_\text{std}$ from \{0.0, 1e-4, 1e-3\}.
We fix $\lambda_\text{match}\text{ = 1e-4 (performance is largely insensitive to this choice)}$, $\tau$ = 1.0, and $C$ = 5. For baseline-specific hyperparameters, we follow the search ranges reported in the original papers.

% For \proposed, the number of interests $N_H, N_C$ is chosen from \{10, 20, ..., 50\}, interest lengths $L_H, L_C$ from \{10, 20, 50\}.
% % neighbor
% The number of similar users $k$ for pseudo-historical knowledge assignment is chosen from \{5, 10, 15, 20, 25\}, and 
% % lambda
% $\lambda_\text{std}, \lambda_\text{match}$ from \{0.0, 1e-4, 1e-3, 1e-2\}.
% $\tau$ are fixed at 1.0, and 
% % cycle
% the update cycles $C_H, C_P$ are both set to 5.
% % temperature
% % $\tau$ is chosen from \{0.5, 1.0, 1.5\},
% % % cycle
% % cycles $C_H, C_P$ from \{1, 5, 20, 100\}, and
% For baseline-specific hyperparameters, we follow the search ranges reported in the original papers.

\vspace{-0.2cm}
\subsection{Performance Comparison}
% We present a comprehensive comparison with baselines as the main results.
\label{sec:result}
% \vspace{-0.1cm}
% \subsubsection{\textbf{Main results.}}
Table \ref{tab:fine_tune} and Table \ref{tab:full_batch} show the overall performance under the fine-tune and full-batch setups, respectively.
In both setups, \proposed consistently outperforms the baselines by effectively retaining historical knowledge (RA) while acquiring current knowledge (LA), achieving a better balance between them (H-mean).
% We provide detailed analyses in both setups as follows.

\vspace{-0.1cm}
\subsubsection{\textbf{Fine-tune setup.}} 
Overall, \proposed shows superior performance across all data blocks compared to state-of-the-art CL methods, including regularization-based (i.e., SAIL-PIW), replay-based (i.e., Reloop2), and multi-interest incremental SR (i.e., IMSR) approaches. 
Unlike conventional CL methods that gradually dilute historical knowledge, CSTRec, equipped with CSA (featuring CSN and CIE), effectively preserves historical user interests (reflected in improved RA) while leveraging them to facilitate the learning of current user interests (reflected in LA). 
Moreover, CSTRec strategically leverages existing users’ historical knowledge to enhance adaptation for new users, further improving overall performance.

\begin{figure}[t!]
\includegraphics[width=0.35\columnwidth]{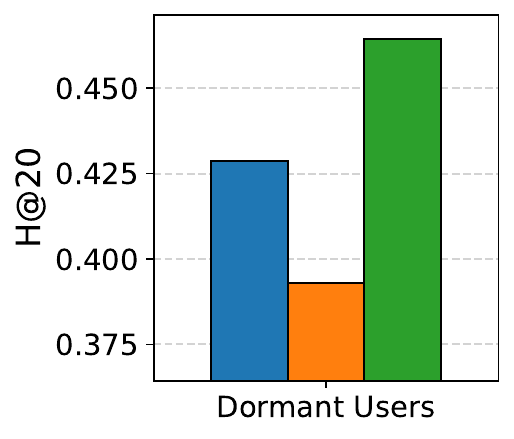}
    \includegraphics[width=0.30\columnwidth]{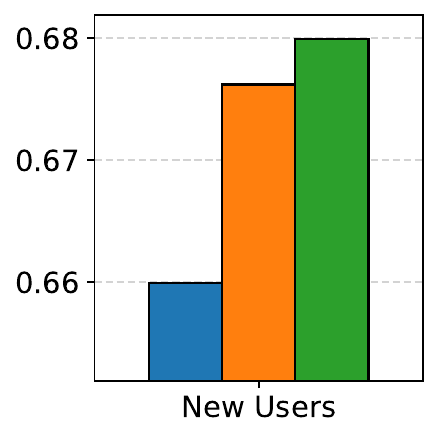}
    \includegraphics[width=0.30\columnwidth]{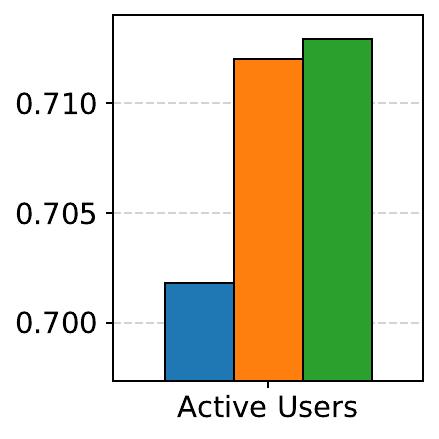}
  \caption{Hit@20 results on Gowalla across three user groups. (Blue: IMSR, Orange: SAIL-PIW, Green: CSTRec)}
  \label{fig:user_study}
  \vspace{-0.3cm}
\end{figure}

\smallsection{Analysis on various user groups.}
For a more thorough assessment of the model's capabilities in (1) knowledge retention, (2) knowledge acquisition, and (3) balancing these aspects, we perform a user-level analysis.
We compare two CL baselines, IMSR and SAIL-PIW, which show competitive results in the main tables.
After fine-tuning on $D_4$, we report the results on three distinct~user~groups:
\begin{enumerate}[leftmargin=*]\vspace{-0.1cm}
    \item \textbf{Dormant users} who interact only in $D_1$ and $D_4$ (i.e., inactive during $D_2$ and $D_3$).
    \item \textbf{New users} who are newly joined in $D_4$.
    \item \textbf{Active users} who interact across all blocks, from $D_1$ to $D_4$.
\end{enumerate}\vspace{-0.1cm}
% results
Figure~\ref{fig:user_study} shows that \proposed consistently outperforms IMSR and SAIL-PIW for all user groups. 
These results collectively support the superiority of \proposed in retaining historical knowledge (dormant users), adapting to current knowledge (new users), and balancing them to provide high-quality recommendations (active users).

\vspace{-0.1cm}
\subsubsection{\textbf{Full-batch setup.}} 
Compared to various long-term SR baselines, including self-attention (i.e., SASRec), and RNN-based (i.e., HPMN), and linear attention (i.e., LimaRec and LinRec) methods, \proposed more effectively captures knowledge from data streams, as evidenced by improved performance across data blocks.
In general, self- and linear attention approaches outperform the RNN-based method by capturing long-range dependencies.
\proposed further enhances performance by capturing the trajectory of user interests over time. 
Specifically, CIE provides additional user-specific guidance by leveraging collaborative signals to enrich both historical and current user interests. 
Moreover, CSN ensures stable learning along data streams, addressing the instability often observed in training linear attention-based methods.
This complementary synergy is reflected in the enhanced $H$-mean performance.
% thereby achieving superior performance over LimaRec and LinRec. 

In summary, \proposed enables effective adaptation to continuously arriving data while mitigating the forgetting of previously acquired knowledge during fine-tuning.
\proposed also shows strong capabilities in capturing long-term preferences from data streams and improving recommendation quality in full-batch setup.
These results collectively support the effectiveness of \proposed in handling continuously incoming user behavior sequences for continual SR.

\begin{table*}[ht!]
    \footnotesize
    \renewcommand{\arraystretch}{0.25}
    \centering
    
    % 첫 번째 테이블
    \vspace{-0.1cm}
    \resizebox{\linewidth}{!}{%
    \begin{minipage}{\linewidth}
        \centering
        \caption{Overall performance comparison for fine-tune setup. * indicates $p < 0.05$ for the paired t-test against the best baseline.}
        % \begin{table*}[ht!]
%     \caption{Overall performance comparison for fine-tune setup. * indicates $p < 0.05$ for the paired t-test against the best baseline.}
    
%     \renewcommand{\arraystretch}{0.2}  % 행 간격을 줄임
%     % \renewcommand{\tabcolsep}{1.2mm}

%     \centering
%     \resizebox{\linewidth}{!}{%
\begin{tabular}{ccc|ccc|ccc|ccc}
        \toprule
        \multicolumn{3}{c|}{\multirow{2}{*}{\textbf{Fine-tune}}}                                                                            & \multicolumn{3}{c|}{\textbf{After $D_2$}}           & \multicolumn{3}{c|}{\textbf{After $D_3$}}           & \multicolumn{3}{c}{\textbf{After $D_4$}}            \\
        \multicolumn{3}{c|}{}                                                                                                               & \textbf{RA}     & \textbf{LA}     & \textbf{H-mean} & \textbf{RA}     & \textbf{LA}     & \textbf{H-mean} & \textbf{RA}     & \textbf{LA}     & \textbf{H-mean} \\ \hline\hline

        % \rotatebox[origin=c]{90}{\textbf{Gowalla}} 
        \multicolumn{1}{c|}{\multirow{43}{*}{\textbf{Gowalla}}} & 
         \multicolumn{1}{c|}{\multirow{12}{*}{\textbf{Hit@20}}} 
         & SASRec              & 0.5157          & 0.6242          & 0.5648          & 0.4456          & 0.6423          & 0.5261          & 0.3890          & 0.6486          & 0.4863          \\
        \multicolumn{1}{c|}{}                                   &\multicolumn{1}{c|}{}                                  & SAIL-PIW            & 0.5153          & 0.6239          & 0.5644          & 0.4476          & {\ul 0.6426}    & 0.5276          & 0.3901          & {\ul 0.6492}    & 0.4874          \\
        \multicolumn{1}{c|}{}                                   &\multicolumn{1}{c|}{}                                  & Reloop2             & 0.5158          & {\ul 0.6243}    & {\ul 0.5649}    & 0.4456          & 0.6423          & 0.5261          & 0.3890          & 0.6486          & 0.4863          \\
        \multicolumn{1}{c|}{}                                   &\multicolumn{1}{c|}{}                                  & IMSR                & {\ul 0.5177}    & 0.6162          & 0.5627          & {\ul 0.4526}    & 0.6350          & {\ul 0.5285}    & {\ul 0.3980}    & 0.6412          & {\ul 0.4912}    \\
        \multicolumn{1}{c|}{}                                   &\multicolumn{1}{c|}{}                                  & \textbf{CSTRec} & \textbf{0.5356*} & \textbf{0.6281*} & \textbf{0.5782*} & \textbf{0.4747*} & \textbf{0.6468*} & \textbf{0.5476*} & \textbf{0.4247*} & \textbf{0.6545*} & \textbf{0.5151*} \\ \cline{2-12}

    \multicolumn{1}{c|}{}                                   & \multicolumn{1}{c|}{\multirow{12}{*}{\textbf{MRR@20}}} &
 SASRec              & 0.3806          & \textbf{0.4661} & 0.4191    & 0.3128          & 0.4724    & 0.3763          & 0.2550          & 0.4728          & 0.3313          \\
        \multicolumn{1}{c|}{}                                   & \multicolumn{1}{c|}{}                               & SAIL-PIW            & 0.3814    & {\ul 0.4652}    & {\ul0.4192} & 0.3150          & {\ul0.4725} & {\ul 0.3780}    & 0.2553          & {\ul 0.4735}    & 0.3318          \\
        \multicolumn{1}{c|}{}                                   & \multicolumn{1}{c|}{}                               & Reloop2             & 0.3806          & \textbf{0.4661} & 0.4191    & 0.3127          & 0.4724    & 0.3763          & 0.2548          & 0.4728          & 0.3312          \\
        \multicolumn{1}{c|}{}                                   & \multicolumn{1}{c|}{}                               & IMSR                & {\ul0.3828} & 0.4498          & 0.4136          & {\ul 0.3187}    & 0.4596          & 0.3764          & {\ul 0.2609}    & 0.4606          & {\ul 0.3332}    \\
        \multicolumn{1}{c|}{}                                   & \multicolumn{1}{c|}{}                               & \textbf{CSTRec} & \textbf{0.3939*}          & 0.4631          & \textbf{0.4257*}          & \textbf{0.3301*} & \textbf{0.4731}          & \textbf{0.3889*} & \textbf{0.2681*} & \textbf{0.4747} & \textbf{0.3423*} \\ \cline{2-12}

        \multicolumn{1}{c|}{}                                   & \multicolumn{1}{c|}{\multirow{12}{*}{\textbf{NDCG@20}}} & SASRec              & 0.4118          & \textbf{0.5029} & {\ul0.4529} & 0.3433          & 0.5120          & 0.4110          & 0.2854          & 0.5139          & 0.3670          \\
        \multicolumn{1}{c|}{}                                   & \multicolumn{1}{c|}{}                               & SAIL-PIW            & 0.4123    & {\ul 0.5022}    &  0.4528    & 0.3455          & {\ul 0.5121}    & {\ul 0.4126}    & 0.2860          & {\ul 0.5145}    & 0.3677          \\
        \multicolumn{1}{c|}{}                                   & \multicolumn{1}{c|}{}                               & Reloop2             & 0.4117          & \textbf{0.5029} & 0.4528    & 0.3433          & 0.5119          & 0.4110          & 0.2853          & 0.5138          & 0.3669          \\
        \multicolumn{1}{c|}{}                                   & \multicolumn{1}{c|}{}                               & IMSR                & {\ul0.4140} & 0.4885          & 0.4482          & {\ul 0.3494}    & 0.5002          & 0.4114          & {\ul 0.2920}    & 0.5026          & {\ul 0.3693}    \\
        \multicolumn{1}{c|}{}                                   & \multicolumn{1}{c|}{}                               & \textbf{CSTRec} & \textbf{0.4265*}          & 0.5015          & \textbf{0.4610*}          & \textbf{0.3631*} & \textbf{0.5135} & \textbf{0.4254*} & \textbf{0.3035*} & \textbf{0.5164} & \textbf{0.3820*} \\ \midrule\midrule
        
        % \multicolumn{1}{c|}{\multirow{41}{*}{\rotatebox[origin=c]{90}{\textbf{ML-1M}} }}   & 

          \multicolumn{1}{c|}{\multirow{41}{*}{\textbf{ML-1M}}} & 

         \multicolumn{1}{c|}{\multirow{12}{*}{\textbf{Hit@20}}}

         & SASRec              & 0.5783          & 0.7710          & 0.6609          & 0.4768          & 0.7594          & 0.5858          & 0.3511          & 0.7119          & 0.4703          \\
        \multicolumn{1}{c|}{}    
        & \multicolumn{1}{c|}{}                                  & SAIL-PIW            & {\ul 0.5807}    & {\ul 0.7726}    & {\ul 0.6630}    & {\ul 0.4780}    & {\ul 0.7608}    & {\ul 0.5872}    & {\ul 0.3525}    & {\ul 0.7136}    & {\ul 0.4719}    \\
        \multicolumn{1}{c|}{}    
        & \multicolumn{1}{c|}{}                                  & Reloop2             & 0.5785          & 0.7714          & 0.6612          & 0.4771          & 0.7597          & 0.5861          & 0.3513          & 0.7122          & 0.4706          \\
        \multicolumn{1}{c|}{}    
        & \multicolumn{1}{c|}{}                                  & IMSR                & 0.5754          & 0.7646          & 0.6566          & 0.4763          & 0.7479          & 0.5820          & 0.3484          & 0.6969          & 0.4646          \\
        \multicolumn{1}{c|}{}    
        & \multicolumn{1}{c|}{}                                  & \textbf{CSTRec} & \textbf{0.5857} & \textbf{0.7815*} & \textbf{0.6696*} & \textbf{0.4830*} & \textbf{0.7714*} & \textbf{0.5940*} & \textbf{0.3555} & \textbf{0.7311*} & \textbf{0.4784*} \\ \cline{2-12}

    \multicolumn{1}{c|}{}                                   & \multicolumn{1}{c|}{\multirow{12}{*}{\textbf{MRR@20}}} &
    SASRec              & 0.1383          & 0.1755          & 0.1547          & 0.1139          & 0.1699          & 0.1363          & 0.0780          & 0.1535          & 0.1034          \\
        \multicolumn{1}{c|}{}                                   & \multicolumn{1}{c|}{}                               & SAIL-PIW            & {\ul 0.1387}    & {\ul 0.1764}    & {\ul 0.1553}    & 0.1143          & {\ul 0.1708}    & {\ul 0.1369}    & {\ul 0.0785}    & {\ul 0.1544}    & {\ul 0.1041}    \\
        \multicolumn{1}{c|}{}                                   & \multicolumn{1}{c|}{}                               & Reloop2             & 0.1384          & 0.1755          & 0.1548          & 0.1136          & 0.1699          & 0.1362          & 0.0781          & 0.1533          & 0.1034          \\
        \multicolumn{1}{c|}{}                                   & \multicolumn{1}{c|}{}                               & IMSR                & 0.1385          & 0.1756          & 0.1549          & {\ul 0.1149}    & 0.1690          & 0.1368          & 0.0778          & 0.1517          & 0.1029          \\
        \multicolumn{1}{c|}{}                                   & \multicolumn{1}{c|}{}                               & \textbf{CSTRec} & \textbf{0.1484*} & \textbf{0.1940*} & \textbf{0.1682*} & \textbf{0.1204*} & \textbf{0.1885*} & \textbf{0.1470*} & \textbf{0.0823*} & \textbf{0.1732*} & \textbf{0.1115*} \\ \cline{2-12} 
        \multicolumn{1}{c|}{}                                   & \multicolumn{1}{c|}{\multirow{12}{*}{\textbf{NDCG@20}}} & SASRec              & 0.2334          & 0.3041          & 0.2641          & 0.1921          & 0.2971          & 0.2333          & 0.1364          & 0.2735          & 0.1821          \\
        \multicolumn{1}{c|}{}                                   & \multicolumn{1}{c|}{}                               & SAIL-PIW            & {\ul 0.2342}    & {\ul 0.3051}    & {\ul 0.2650}    & 0.1928          & {\ul 0.2980}    & {\ul 0.2341}    & {\ul 0.1371}    & {\ul 0.2746}    & {\ul 0.1829}    \\
        \multicolumn{1}{c|}{}                                   & \multicolumn{1}{c|}{}                               & Reloop2             & 0.2334          & 0.3042          & 0.2642          & 0.1920          & 0.2971          & 0.2332          & 0.1365          & 0.2733          & 0.1821          \\
        \multicolumn{1}{c|}{}                                   & \multicolumn{1}{c|}{}                               & IMSR                & 0.2329          & 0.3028          & 0.2633          & {\ul 0.1929}    & 0.2938          & 0.2328          & 0.1357          & 0.2688          & 0.1804          \\
        \multicolumn{1}{c|}{}                                   & \multicolumn{1}{c|}{}                               & \textbf{CSTRec} & \textbf{0.2434*} & \textbf{0.3218*} & \textbf{0.2772*} & \textbf{0.1990*} & \textbf{0.3152*} & \textbf{0.2440*} & \textbf{0.1410*} & \textbf{0.2940*} & \textbf{0.1906*} \\ \midrule\midrule

        % yelp
        
          \multicolumn{1}{c|}{\multirow{44}{*}{\textbf{Yelp}}} & 

         \multicolumn{1}{c|}{\multirow{12}{*}{\textbf{Hit@20}}} & SASRec   & 0.1076          & 0.1220           & 0.1143          & 0.0985          & 0.1242          & 0.1098          & 0.0886           & 0.1263          & 0.1041          \\
        \multicolumn{1}{c|}{}  & \multicolumn{1}{c|}{}                                   & SAIL-PIW & {\ul 0.1108}    & {\ul 0.1249}     & {\ul 0.1174}    & {\ul 0.1001}    & {\ul 0.1320}    & {\ul 0.1138}    & 0.0873           & {\ul 0.1349}    & {\ul 0.1059} \\
                                 \multicolumn{1}{c|}{}                                   & \multicolumn{1}{c|}{}                                   & Reloop2                      & 0.1076          & 0.1219          & 0.1143          & 0.0985          & 0.1243          & 0.1099          & 0.0886    & 0.1263          & 0.1042          \\
                                 \multicolumn{1}{c|}{}                                   & \multicolumn{1}{c|}{}                                   & IMSR                         & 0.1075          & 0.1206          & 0.1137          & 0.0973          & 0.1232          & 0.1087          & {\ul 0.0890} & 0.1253          & 0.1042          \\
                                 \multicolumn{1}{c|}{}                                   & \multicolumn{1}{c|}{}  & \textbf{CSTRec} & \textbf{0.1175*} & \textbf{0.1383*}  & \textbf{0.1271*} & \textbf{0.1048*} & \textbf{0.1427*} & \textbf{0.1208*} & \textbf{0.0895} & \textbf{0.1431*} & \textbf{0.1101*} \\
    \\ \cline{2-12} 
        \multicolumn{1}{c|}{}                                   & \multicolumn{1}{c|}{\multirow{12}{*}{\textbf{MRR@20}}}
                                                              
        & SASRec                       & 0.0232          & 0.0251          & 0.0241          & 0.0210          & 0.0257          & 0.0231          & {\ul0.0189} & 0.0260          & 0.0219          \\
                                 \multicolumn{1}{c|}{}                                   & \multicolumn{1}{c|}{}                                  & SAIL-PIW & {\ul 0.0243}    & {\ul 0.0260}     & {\ul 0.0251}    & {\ul 0.0211}    & {\ul 0.0271}    & {\ul 0.0237}    & 0.0185           & {\ul 0.0278}    & {\ul 0.0222}    \\
                                 \multicolumn{1}{c|}{}                                   & \multicolumn{1}{c|}{}                                  & Reloop2                      & 0.0232          & 0.0251          & 0.0241          & 0.0210          & 0.0257          & 0.0231          & {\ul 0.0189} & 0.0260          & 0.0219          \\
                                 \multicolumn{1}{c|}{}                                   & \multicolumn{1}{c|}{}                                  & IMSR                         & 0.0236    & 0.0253          & 0.0244          & 0.0208          & 0.0258          & 0.0230          & 0.0187    & 0.0260          & 0.0217          \\

                                 \multicolumn{1}{c|}{}                                   & \multicolumn{1}{c|}{}  & \textbf{CSTRec}   & \textbf{0.0250} & \textbf{0.0282*} & \textbf{0.0265*} & \textbf{0.0216} & \textbf{0.0287*} & \textbf{0.0246*} & \textbf{0.0190}  & \textbf{0.0287} & \textbf{0.0226} \\ \cline{2-12} 
                                 \multicolumn{1}{c|}{}                                   & \multicolumn{1}{c|}{\multirow{12}{*}{\textbf{NDCG@20}}}
                                 & SASRec                       & 0.0411          & 0.0457          & 0.0433          & 0.0375          & 0.0466          & 0.0415          & {\ul 0.0337} & 0.0473          & 0.0394          \\
                                 \multicolumn{1}{c|}{}                                   & \multicolumn{1}{c|}{}                                  & SAIL-PIW & {\ul 0.0426}    & {\ul 0.0470}     & {\ul 0.0448}    & {\ul 0.0378}    & {\ul 0.0494}    & {\ul 0.0428}    & 0.0331           & {\ul 0.0505}    & {\ul 0.0400}    \\
                                 \multicolumn{1}{c|}{}                                   & \multicolumn{1}{c|}{}                                  & Reloop2                      & 0.0411          & 0.0457          & 0.0433          & 0.0375          & 0.0466          & 0.0415          & {\ul 0.0337} & 0.0473          & 0.0394          \\
                                 \multicolumn{1}{c|}{}                                   & \multicolumn{1}{c|}{}                                  & IMSR                         & 0.0414          & 0.0455          & 0.0434          & 0.0371          & 0.0464          & 0.0412          & 0.0336    & 0.0471          & 0.0392          \\
                                 \multicolumn{1}{c|}{}                                   & \multicolumn{1}{c|}{} 
& \textbf{CSTRec}   & \textbf{0.0446*} & \textbf{0.0515*}  & \textbf{0.0479*} & \textbf{0.0393*} & \textbf{0.0528*} & \textbf{0.0451*} & \textbf{0.0338}  & \textbf{0.0529*} & \textbf{0.0412*} \\ \bottomrule
        \end{tabular}
%         }
%     \label{tab:main_FT_one}
% \end{table*}
  % 첫 번째 테이블
        \label{tab:fine_tune}  % 첫 번째 테이블에 대한 label
    \end{minipage}
    }
    \vspace{0.1cm}  % 두 테이블 사이에 간격 추가

    % 두 번째 테이블
    \resizebox{\linewidth}{!}{%
    \begin{minipage}{\linewidth}
        \centering
        \caption{Overall performance comparison for full-batch setup. * indicates $p < 0.05$ for the paired t-test against the best baseline.}
        % \begin{table*}[t!]
%     \caption{Overall performance comparison for full-batch setup. * indicates $p < 0.05$ for the paired t-test against the best baseline.}
%     \renewcommand{\arraystretch}{0.4}  % 행 간격을 줄임

%     \centering
%     \resizebox{\linewidth}{!}{%
\begin{tabular}{ccc|ccc|ccc|ccc}
        \toprule
        \multicolumn{3}{c|}{\multirow{1}{*}{\textbf{Full-batch}}}                                                                            & \multicolumn{3}{c|}{\textbf{After $D_2$}}           & \multicolumn{3}{c|}{\textbf{After $D_3$}}           & \multicolumn{3}{c}{\textbf{After $D_4$}}            \\
        \multicolumn{3}{c|}{}                                                                                                               & \textbf{RA}     & \textbf{LA}     & \textbf{H-mean} & \textbf{RA}     & \textbf{LA}     & \textbf{H-mean} & \textbf{RA}     & \textbf{LA}     & \textbf{H-mean} \\ \hline\hline

        % \multicolumn{1}{c|}{\multirow{15}{*}{\rotatebox[origin=c]{90}{\textbf{Gowalla}} }} & 
          \multicolumn{1}{c|}{\multirow{41}{*}{\textbf{Gowalla}}} & 
        
         \multicolumn{1}{c|}{\multirow{12}{*}{\textbf{Hit@20}}}  
        & SASRec              & 0.7125          & {\ul 0.7117} & 0.7121          & {\ul 0.7190}    & 0.7172          & {\ul 0.7181}    & {\ul 0.7155}    & {\ul 0.7137}    & {\ul 0.7146}    \\
        \multicolumn{1}{c|}{}                                  &\multicolumn{1}{c|}{}                                  & HPMN                & 0.6916          & 0.6886          & 0.6901          & 0.6991          & 0.6939          & 0.6965          & 0.6973          & 0.6914          & 0.6943          \\
        \multicolumn{1}{c|}{}                                  &\multicolumn{1}{c|}{}                                  & LimaRec             & 0.6537          & 0.6286          & 0.6409          & 0.6743          & 0.6333          & 0.6532          & 0.6848          & 0.6354          & 0.6592          \\
        \multicolumn{1}{c|}{}                                  &\multicolumn{1}{c|}{}                                  & LinRec              & {\ul 0.7132}    & {\ul 0.7117} & {\ul 0.7125}    & 0.7169          & {\ul 0.7159}    & 0.7164          & 0.7138          & 0.7128          & 0.7133          \\
        \multicolumn{1}{c|}{}                                  &\multicolumn{1}{c|}{}                                  & \textbf{CSTRec} & \textbf{0.7200*} & \textbf{0.7177*} & \textbf{0.7189*} & \textbf{0.7263*} & \textbf{0.7230*} & \textbf{0.7246*} & \textbf{0.7218*} & \textbf{0.7188*} & \textbf{0.7203*} \\ \cline{2-12}

    \multicolumn{1}{c|}{}                                   & \multicolumn{1}{c|}{\multirow{12}{*}{\textbf{MRR@20}}} &
 SASRec              & 0.5906          & 0.5732          & 0.5818          & 0.5938          & 0.5703          & 0.5818          & \textbf{ 0.5850}    & 0.5690          & 0.5769          \\
            \multicolumn{1}{c|}{}                                   & \multicolumn{1}{c|}{}                               & HPMN                & 0.5846          & 0.5745          & 0.5795          & 0.5673          & 0.5587          & 0.5629          & 0.5625          & 0.5554          & 0.5590          \\
            \multicolumn{1}{c|}{}                                   & \multicolumn{1}{c|}{}                               & LimaRec             & 0.4788          & 0.3978          & 0.4338          & 0.4714          & 0.3846          & 0.4231          & 0.5062          & 0.3949          & 0.4434          \\
            \multicolumn{1}{c|}{}                                   & \multicolumn{1}{c|}{}                               & LinRec              & \textbf{0.6080} & {\ul 0.5881}    & {\ul 0.5979}    & {\ul 0.6026}    & {\ul 0.5832}    & {\ul 0.5927}    & 0.5819 & \textbf{0.5759} & \textbf{0.5789} \\
            \multicolumn{1}{c|}{}                                   & \multicolumn{1}{c|}{}                               & \textbf{CSTRec} & {\ul 0.6077}    & \textbf{0.5920} & \textbf{0.5997} & \textbf{0.6054*} & \textbf{0.5867} & \textbf{0.5958*} & {\ul 0.5834}          & {\ul 0.5731}    & {\ul 0.5782}    \\ \cline{2-12} 
            \multicolumn{1}{c|}{}                                   & \multicolumn{1}{c|}{\multirow{12}{*}{\textbf{NDCG@20}}} & SASRec              & 0.6190          & 0.6056          & 0.6122          & 0.6231          & 0.6048          & 0.6138          & {\ul 0.6156}          & 0.6029          & 0.6092          \\
            \multicolumn{1}{c|}{}                                   & \multicolumn{1}{c|}{}                               & HPMN                & 0.6095          & 0.6011          & 0.6053          & 0.5981          & 0.5904          & 0.5942          & 0.5940          & 0.5873          & 0.5906          \\
            \multicolumn{1}{c|}{}                                   & \multicolumn{1}{c|}{}                               & LimaRec             & 0.5191          & 0.4505          & 0.4820          & 0.5187          & 0.4416          & 0.4767          & 0.5479          & 0.4501          & 0.4941          \\
            \multicolumn{1}{c|}{}                                   & \multicolumn{1}{c|}{}                               & LinRec              & {\ul 0.6326}    & {\ul 0.6170}    & {\ul 0.6247}    & {\ul 0.6292}    & {\ul 0.6143}    & {\ul 0.6217}    & 0.6119 & {\ul0.6072} & {\ul 0.6096} \\
            \multicolumn{1}{c|}{}                                   & \multicolumn{1}{c|}{}                               & \textbf{CSTRec} & \textbf{0.6339*} & \textbf{0.6214} & \textbf{0.6276} & \textbf{0.6335*} & \textbf{0.6186*} & \textbf{0.6259*} & \textbf{ 0.6158*}    & \textbf{ 0.6073}    & \textbf{ 0.6115}   \\ \midrule\midrule
        
        % \multicolumn{1}{c|}{\multirow{15}{*}{\rotatebox[origin=c]{90}{\textbf{ML-1M}} }}   & 
        \multicolumn{1}{c|}{\multirow{40}{*}{\textbf{ML-1M}}} & 

         \multicolumn{1}{c|}{\multirow{12}{*}{\textbf{Hit@20}}}

        & SASRec              & {\ul 0.6669}    & {\ul 0.7208}    & {\ul 0.6928}    & {\ul 0.5598}    & {\ul 0.6485}    & {\ul 0.6009}    & {\ul 0.4559}    & {\ul 0.5676}    & {\ul 0.5057}    \\
        \multicolumn{1}{c|}{}                                  & \multicolumn{1}{c|}{}                                  & HPMN                & 0.1321          & 0.1387          & 0.1353          & 0.1229          & 0.1301          & 0.1264          & 0.1082          & 0.1184          & 0.1131          \\
        \multicolumn{1}{c|}{}                                  & \multicolumn{1}{c|}{}                                  & LimaRec             & 0.4554          & 0.4919          & 0.4728          & 0.3959          & 0.4418          & 0.4174          & 0.3411          & 0.3859          & 0.3619          \\
        \multicolumn{1}{c|}{}                                  & \multicolumn{1}{c|}{}                                  & LinRec              & 0.4783          & 0.5314          & 0.5035          & 0.4168          & 0.4807          & 0.4465          & 0.3446          & 0.4194          & 0.3783          \\
        \multicolumn{1}{c|}{}                                  & \multicolumn{1}{c|}{}                                  & \textbf{CSTRec} & \textbf{0.6764*} & \textbf{0.7248} & \textbf{0.6998*} & \textbf{0.5799*} & \textbf{0.6548*} & \textbf{0.6151*} & \textbf{0.4785*} & \textbf{0.5776*} & \textbf{0.5234*}  \\ \cline{2-12} 
        
    \multicolumn{1}{c|}{}                                   & \multicolumn{1}{c|}{\multirow{12}{*}{\textbf{MRR@20}}} &
   SASRec              & {\ul 0.1542}    & {\ul 0.1738}    & {\ul 0.1635}    & {\ul 0.1295}    & {\ul 0.1536}    & {\ul 0.1405}    & {\ul 0.1063}    & {\ul 0.1321}    & {\ul 0.1178}    \\
            \multicolumn{1}{c|}{}                                   & \multicolumn{1}{c|}{}                               & HPMN                & 0.0229          & 0.0239          & 0.0234          & 0.0219          & 0.0227          & 0.0223          & 0.0197          & 0.0210          & 0.0203          \\
            \multicolumn{1}{c|}{}                                   & \multicolumn{1}{c|}{}                               & LimaRec             & 0.1084          & 0.1160          & 0.1120          & 0.0966          & 0.1050          & 0.1005          & 0.0830          & 0.0912          & 0.0869          \\
            \multicolumn{1}{c|}{}                                   & \multicolumn{1}{c|}{}                               & LinRec              & 0.1206          & 0.1336          & 0.1268          & 0.1052          & 0.1201          & 0.1122          & 0.0865          & 0.1040          & 0.0945          \\
            \multicolumn{1}{c|}{}                                   & \multicolumn{1}{c|}{}                               & \textbf{CSTRec} & \textbf{0.1658*} & \textbf{0.1901*} & \textbf{0.1771*} & \textbf{0.1381*} & \textbf{0.1660*} & \textbf{0.1508*} & \textbf{0.1123*} & \textbf{0.1430*} & \textbf{0.1258*} \\ \cline{2-12}
            \multicolumn{1}{c|}{}                                   & \multicolumn{1}{c|}{\multirow{12}{*}{\textbf{NDCG@20}}} & SASRec              & {\ul 0.2646}    & {\ul 0.2925}    & {\ul 0.2779}    & {\ul 0.2216}    & {\ul 0.2604}    & {\ul 0.2395}    & {\ul 0.1811}    & {\ul 0.2259}    & {\ul 0.2010}    \\
            \multicolumn{1}{c|}{}                                   & \multicolumn{1}{c|}{}                               & HPMN                & 0.0458          & 0.0479          & 0.0468          & 0.0431          & 0.0452          & 0.0441          & 0.0383          & 0.0414          & 0.0398          \\
            \multicolumn{1}{c|}{}                                   & \multicolumn{1}{c|}{}                               & LimaRec             & 0.1828          & 0.1968          & 0.1895          & 0.1609          & 0.1774          & 0.1686          & 0.1384          & 0.1544          & 0.1459          \\
            \multicolumn{1}{c|}{}                                   & \multicolumn{1}{c|}{}                               & LinRec              & 0.1979          & 0.2196          & 0.2082          & 0.1724          & 0.1980          & 0.1843          & 0.1421          & 0.1720          & 0.1557          \\
            \multicolumn{1}{c|}{}                                   & \multicolumn{1}{c|}{}                               & \textbf{CSTRec} & \textbf{0.2764*} & \textbf{0.3069*} & \textbf{0.2909*} & \textbf{0.2332*} & \textbf{0.2722*} & \textbf{0.2512*} & \textbf{0.1909*} & \textbf{0.2371*} & \textbf{0.2115*} 
        \\ 
        \midrule\midrule
        % yelp

 \multicolumn{1}{c|}{\multirow{41}{*}{\textbf{Yelp}}} & 

         \multicolumn{1}{c|}{\multirow{12}{*}{\textbf{Hit@20}}} 
                        & SASRec                       & 0.1012                & 0.0981          & 0.0997          & 0.1098                & 0.0986          & 0.1039          & 0.1144          & 0.1012          & 0.1074                \\
                                 \multicolumn{1}{c|}{}                                   & \multicolumn{1}{c|}{}                                  & HPMN                         & {\ul 0.1537}          & {\ul 0.1462}    & {\ul 0.1499}    & {\ul 0.1438} & {\ul 0.1436}    & {\ul 0.1437}    & \textbf{0.1299}       & {\ul 0.1389}    & {\ul 0.1343} \\
                                 \multicolumn{1}{c|}{}                                   & \multicolumn{1}{c|}{}                                  & LimaRec                      & 0.1018                & 0.1010          & 0.1014          & 0.0928                & 0.0930          & 0.0929          & 0.0864          & 0.0868          & 0.0865                \\
                                 \multicolumn{1}{c|}{}                                   & \multicolumn{1}{c|}{}                                  & LinRec                       & 0.1440                & 0.1299          & 0.1366          & 0.1388                & 0.1286          & 0.1335          & 0.1289 & 0.1259          & 0.1274                \\
                                 \multicolumn{1}{c|}{}                                   & \multicolumn{1}{c|}{} &
                               \textbf{CSTRec} & \textbf{0.1647*}       & \textbf{0.1605*} & \textbf{0.1626*} & \textbf{0.1486*} & \textbf{0.1529*} & \textbf{0.1507*} & {\ul0.1290}                & \textbf{0.1425*} &  \textbf{0.1353} \\ \cline{2-12} 
    \multicolumn{1}{c|}{}                                   & \multicolumn{1}{c|}{\multirow{12}{*}{\textbf{MRR@20}}}                       & SASRec                       & 0.0221                & 0.0214          & 0.0218          & 0.0238                & 0.0210          & 0.0223          & 0.0249       & 0.0218          & 0.0233                \\
                                 \multicolumn{1}{c|}{}                                   & \multicolumn{1}{c|}{}                                 & HPMN                         & {\ul 0.0321} & {\ul 0.0301}    & {\ul 0.0311}    & {\ul 0.0300}          & {\ul 0.0294}    & {\ul 0.0297}    & 0.0274          & {\ul 0.0284}    & {\ul 0.0279} \\
                                 \multicolumn{1}{c|}{}                                   & \multicolumn{1}{c|}{}                                 & LimaRec                      & 0.0220                & 0.0217          & 0.0219          & 0.0203                & 0.0199          & 0.0201          & 0.0185       & 0.0185          & 0.0185                \\
                                 \multicolumn{1}{c|}{}                                   & \multicolumn{1}{c|}{}                                 & LinRec                       & 0.0311          & 0.0278          & 0.0294          & 0.0295                & 0.0272          & 0.0283          &  {\ul 0.0275}          & 0.0267          & 0.0271                \\
                                 \multicolumn{1}{c|}{}                                   & \multicolumn{1}{c|}{} & \textbf{CSTRec}                         & \textbf{0.0343*}       & \textbf{0.0325*} & \textbf{0.0334*} & \textbf{0.0308*}       & \textbf{0.0307*} & \textbf{0.0307*} & \textbf{0.0279}       & \textbf{0.0289} & \textbf{0.0284} \\ \cline{2-12} 
                                     \multicolumn{1}{c|}{}                                   & \multicolumn{1}{c|}{\multirow{12}{*}{\textbf{NDCG@20}}}                                  & SASRec                       & 0.0390                & 0.0378          & 0.0384          & 0.0420                & 0.0376          & 0.0397          & 0.0439       & 0.0387          & 0.0411                \\
                                 \multicolumn{1}{c|}{}                                   & \multicolumn{1}{c|}{}                                 & HPMN                         & {\ul 0.0580}          & {\ul 0.0547}    & {\ul 0.0563}    & {\ul 0.0541} & {\ul 0.0536}    & {\ul 0.0539}    & {\ul 0.0492}          & {\ul 0.0518}    & {\ul 0.0505} \\
                                 \multicolumn{1}{c|}{}                                   & \multicolumn{1}{c|}{}                                 & LimaRec                      & 0.0390                & 0.0385          & 0.0388          & 0.0357                & 0.0354          & 0.0356          & 0.0329       & 0.0330          & 0.0330                \\
                                 \multicolumn{1}{c|}{}                                   & \multicolumn{1}{c|}{}                                 & LinRec                       & 0.0551                & 0.0495          & 0.0522          & 0.0527                & 0.0487          & 0.0506          & 0.0491          & 0.0477          & 0.0484                \\
\multicolumn{1}{c|}{}                                   & \multicolumn{1}{c|}{} & \textbf{CSTRec}                         & \textbf{0.0620*}        & \textbf{0.0597*} & \textbf{0.0608*} & \textbf{0.0558*} & \textbf{0.0565*} & \textbf{0.0561*} & \textbf{0.0494}       & \textbf{0.0529*} & \textbf{0.0511} \\

        \bottomrule
        \end{tabular}
%         }

%     \label{tab:main_FB_one}
% \end{table*}
  % 두 번째 테이블
        \label{tab:full_batch}  % 두 번째 테이블에 대한 label
    \end{minipage}
    }
    \vspace{-0.3cm}
\end{table*}

%\input{sections/abl_attention}
% \usepackage[normalem]{ulem}
% \begin{table}[t!]
% \caption{Accuracy and efficiency of attention mechanisms}
% \label{tab:att}
% \renewcommand{\arraystretch}{0.75}
% \resizebox{\linewidth}{!}{%
% \begin{tabular}{l|ccccc}
% \toprule
% \noalign{\vskip 0.5mm}
% \textbf{Method} & \textbf{H@20} & \textbf{M@20} & \textbf{N@20} & \textbf{\begin{tabular}[c]{@{}c@{}}Training\\time (s)\end{tabular}} & \textbf{\begin{tabular}[c]{@{}c@{}}Inference\\time (s)\end{tabular}} \\ \midrule\midrule
% SASRec (self-attention)    & 0.7162          & 0.5828          & 0.6142          & 225.26                   & 4.57                        \\
% LimaRec (linear attention) & 0.6730          & 0.4754          & 0.5214          & 249.43                   & 3.46                        \\
% CSTRec (CSN only)          & {\ul 0.7168}    & {\ul 0.5888}    & {\ul 0.6188}    & \textbf{177.70}                   & \textbf{1.97}               \\
% CSTRec (Full CSA)          & \textbf{0.7218} & \textbf{0.5965} & \textbf{0.6259} & {\ul 216.12}                   & {\ul 2.44}                  \\ \bottomrule
% \end{tabular}
% }
% \end{table}

\begin{table}[t!]
\caption{Accuracy and efficiency of attention mechanisms.}
\label{tab:att}
\renewcommand{\arraystretch}{0.75}
\resizebox{\linewidth}{!}{%
\begin{tabular}{l|ccccc}
\toprule
\noalign{\vskip 0.5mm}
\textbf{Method} & \textbf{H@20} & \textbf{M@20} & \textbf{N@20} & \textbf{\begin{tabular}[c]{@{}c@{}}Training\\time (s)\end{tabular}} & \textbf{\begin{tabular}[c]{@{}c@{}}Inference\\time (s)\end{tabular}} \\ \midrule\midrule
Self-attention (SASRec)    & 0.7162          & 0.5828          & 0.6142          & 225.26                   & 4.57                        \\
Linear attention (LimaRec) & 0.6730          & 0.4754          & 0.5214          & 249.43                   & 3.46                        \\
Linear attention (LinRec) & 0.7102          & 0.5847          & 0.6143          & {\ul 213.90}                   & 2.74                        \\
CSA (CSN only)          & {\ul 0.7168}    & {\ul 0.5888}    & {\ul 0.6188}    & \textbf{177.70}                   & \textbf{1.97}               \\
CSA (CSTRec)          & \textbf{0.7218} & \textbf{0.5965} & \textbf{0.6259} & 216.12                   & {\ul 2.44}                  \\ \bottomrule
\end{tabular}
}\vspace{-0.4cm}
\end{table}
\begin{table}[t!]
\centering
% \caption{Impact of each component trained on $D_4$ of Gowalla.}
\caption{Ablation study on Gowalla under the fine-tine setup.}
\renewcommand{\arraystretch}{0.75}
\renewcommand{\tabcolsep}{2mm}
\resizebox{1.0\linewidth}{!}{%
\begin{tabular}{cccc|cccc}
\toprule
\textbf{CSN} & \textbf{CIE-H} & \textbf{CIE-C} & \textbf{PKA} & \textbf{RA}     & \textbf{LA}     & \textbf{H-mean} & \textbf{\begin{tabular}[c]{@{}c@{}}H-mean\\ Imp (\%)\end{tabular}} \\ \midrule\midrule
x   & x     & x     & x   & 0.3847 & 0.6307 & 0.4779 & 0.00                                                      \\ \midrule
x   & o     & o     & o   & 0.3908 & 0.6321 & 0.4830 & +1.07                                                      \\
o   & x     & o     & o   & 0.4142 & \textbf{0.6550} & 0.5075  & +6.19                                                      \\
o   & o     & x     & o   & {\ul 0.4166} & 0.6533 & {\ul0.5088} & +{\ul6.47}                                                      \\
o   & o     & o     & x   & 0.4074 & 0.6517 & 0.5014 & +4.50                                                      \\
o   & o     & o     & o   & \textbf{0.4247} & {\ul0.6545} & \textbf{0.5151} & +\textbf{7.78}                                                      \\ \bottomrule
\end{tabular}
}
\label{tab:abl_component}
\vspace{-0.4cm}
\end{table}
% \usepackage[normalem]{ulem}

\begin{table}[t!]
\centering
%\caption{Impact of top-$K$ users in PKA on new-user performance.}
\caption{New-user performance with top-$K$ users in PKA.}
\resizebox{1.0\linewidth}{!}{%
\begin{tabular}{c|ccccccc}
\toprule
$\mathbf{K}$    & \textbf{0} (no PKA)      & \textbf{5}            & \textbf{10}     & \textbf{15}     & \textbf{20}           & \textbf{25}              & \multicolumn{1}{l}{\textbf{Imp (\%)}} \\ \hline\hline
\textbf{H@20}  & 0.6677 & {\ul 0.6731} & 0.6711 & 0.6714 & {\ul 0.6731} & \textbf{0.6741} & +0.96                          \\
\textbf{M@20}  & 0.4181 & {\ul 0.4231} & 0.4214 & 0.4192 & 0.4193       & \textbf{0.4305} & +2.97                          \\
\textbf{N@20} & 0.4766 & {\ul 0.4818} & 0.4799 & 0.4786 & 0.4786       & \textbf{0.4877} & +2.33                         \\ \bottomrule
\end{tabular}
}
\vspace{0cm}
\label{tab:abl_topk}
\end{table}

% We compare three attention mechanisms: SASRec (self-attention), LimaRec (linear-attention), and \proposed (CSA) by evaluating their accuracy and efficiency on $D_4$ of Gowalla under a full-batch setup.
%

%\vspace{-0.1cm}
\subsection{Study of \proposed}
\label{subsec:exp_study}
We provide comprehensive analyses of \proposed. In this section, we report the results on Gowalla dataset.

\vspace{-0.1cm}
\subsubsection{\textbf{Accuracy and efficiency analysis.}} 
We compare the accuracy and efficiency of three attention mechanisms: self-attention (SASRec), linear attention (LimaRec, LinRec), and CSA (\proposed).
For CSA, we also compare `CSN only', which excludes all other components, to verify CSN’s standalone efficacy.
Table~\ref{tab:att} presents the results on $D_4$ under a full-batch setup.
Here, training time indicates the total time to complete the training process, and inference time measures the time to generate recommendations for all users.

%\footnote{For linear attention, we follow the implementation of LimaRec instead of LinRec, to compare with a direct application of linear attention excluding additional techniques.}
% For CSA, we compare two versions: `CSN only' (excluding all other components) to demonstrate CSN’s standalone efficacy, and `Full CSA' to assess the combined impact of all modules. 
% 분석
% We observe that \proposed outperforms both SASRec and LimaRec in terms of accuracy and efficiency. Below, we provide a detailed analysis of each aspect.
\vspace{-0.1cm}
\smallsection{Accuracy aspect.}
CSA (CSN only) shows significant performance gains over linear attention (LimaRec), highlighting the effectiveness of CSN for stable optimization of linear attention in the continual SR problem. 
Moreover, CSA (\proposed) further improves accuracy through CIE, which enriches historical and current interests in a complementary manner, as also evidenced by its superior performance over self-attention (SASRec) and linear attention (LinRec).

\smallsection{Efficiency aspect.}
Compared to the self-attention, CSA (\proposed) shows comparable efficiency for training and greatly reduced efficiency for inference.
Building upon the linear attention, CSA maintains the linear complexity with respect to the input length.
Also, the additional computation introduced by the proposed modules is negligible compared to the baseline attention mechanisms. 
In contrast, self-attention (SASRec) incurs the highest inference time due to its quadratic complexity.
Linear attention (LinRec) incurs relatively high inference costs due to the repeated application of its dual-side normalization techniques, while Linear attention (LimaRec) shows the slowest training time because its multi-interest module for all users adds extra computational overhead.
% CSA (CSN only) shows the fastest training and inference times with linear complexity. 
% CSA (\proposed) shows the second best efficiency, as the additional computation introduced by the proposed modules is negligible compared to the baseline attention mechanisms. 

% \vspace{-0.15cm}
\subsubsection{\textbf{Ablation study.}}
% Table~\ref{tab:abl_component} shows the impact of each proposed component under the fine-tune setup.
Table~\ref{tab:abl_component} presents a detailed ablation study on $D_4$ of Gowalla. % under the fine-tune setup.
CIE-H and CIE-C denote the use of historical and current interests, respectively (\cref{sub:CIE}). 
PKA refers to pseudo-historical knowledge assignment (\cref{sub:pseudo_h_new_users}). 
First, CSN contributes to the stable accumulation of knowledge, as evidenced by the overall performance gains when it is included.
Second, both CIE-H and CIE-C effectively enrich historical and current interests, respectively. This is supported by a decrease in RA and LA when each component is excluded.
Third, PKA supports new-user adaptation by leveraging existing historical knowledge. Without it, both RA and LA drop, with a larger decline in RA. 
This suggests that failing to properly accommodate new users can negatively affect overall performance. Performance on new users is analyzed in the next section.
Finally, the best H-mean is achieved when all modules are used together, highlighting the synergy among the proposed components.

% Overall, each component enhances either historical or current knowledge. 
% We interpret the results as follows:
% Third, PKA aids in the adaptation of new users by leveraging existing historical knowledge.
% Without it, both RA and LA decline, underscoring the importance of strategically reusing past insights.

% Therefore, these results validate the effectiveness of our proposed modules and the benefits of their integration.

%\input{sections/abl_corr}

\vspace{-0.15cm}
\subsubsection{\textbf{Hyperparameter study.}}
\label{subsub:hyper}
% \subsubsection{\textbf{Top-$K$ users impact on pseudo-historical knowledge.}}
Table~\ref{tab:abl_topk} presents the impact of top-$K$ users in PKA on recommendation performance for new users.
Here, $K$=0 indicates the case that PKA is not used.
For all values of $K$, PKA shows better performance compared to not using it.
This result indicates that PKA effectively addresses the lack of information in newly incoming users, thereby facilitating their adaptation. 

Figure \ref{fig:_interest_pool} provides results with varying the number ($N$) and length ($L$) of interest pools, where $N$ improves diversity and $L$ adds depth.
We report the averaged H-mean from $D_2$ to $D_4$. 
%We observe that the best performance is achieved when assigning more parameters to the current interest pool compared to the historical interest pool.
We find that allocating more capacity to the current interest pool yields the best results, highlighting the importance of capturing complex and diverse trends in incoming data blocks.
In \proposed, historical knowledge is leveraged to facilitate learning new interests, so fewer parameters suffice.
Also, long-term user preferences encoded in historical pool tend to be less dynamic compared to the variety of transient interests in the current data block, which may also explain this tendency.
Similar trends are also observed across other datasets.
% \begin{figure}[h]
%     \centering
%     % Subfigure for dormant user
%     \begin{subfigure}{0.2\textwidth}
%         \centering
%         \includegraphics[width=\textwidth]{images/dormant_user.pdf}
%         \caption{Dormant User}
%         \label{fig:dormant}
%     \end{subfigure}
%     % Subfigure for new user
%     \begin{subfigure}{0.2\textwidth}
%         \centering
%         \includegraphics[width=\textwidth]{images/new_user.pdf}
%         \caption{New User}
%         \label{fig:new}
%     \end{subfigure}
%     % Subfigure for active user
%     \begin{subfigure}{0.2\textwidth}
%         \centering
%         \includegraphics[width=\textwidth]{images/active_user.pdf}
%         \caption{Active User}
%         \label{fig:active}
%     \end{subfigure}
%     \caption{Comparison of Dormant, New, and Active Users}
%     \label{fig:combined}
% \end{figure}

% \begin{figure}[t!]
% \includegraphics[width=0.35\columnwidth]{images/dormant_user.pdf}
%   \hfill
%     \includegraphics[width=0.298\columnwidth]{images/new_user_new.pdf}
%     \hfill
%     \includegraphics[width=0.315\columnwidth]{images/active_user.pdf}
%   \caption{H@20 performance across three user groups.}
%   \label{fig:gap_images}
%   %\vspace{+0.1cm}
% \end{figure}

\begin{figure}[t]
  \includegraphics[width=0.495\columnwidth]{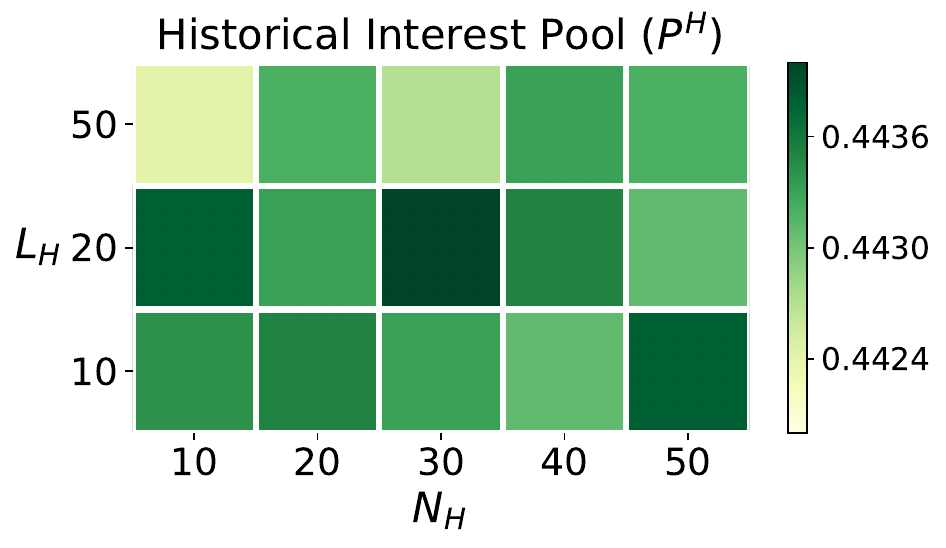}
  \includegraphics[width=0.495\columnwidth]{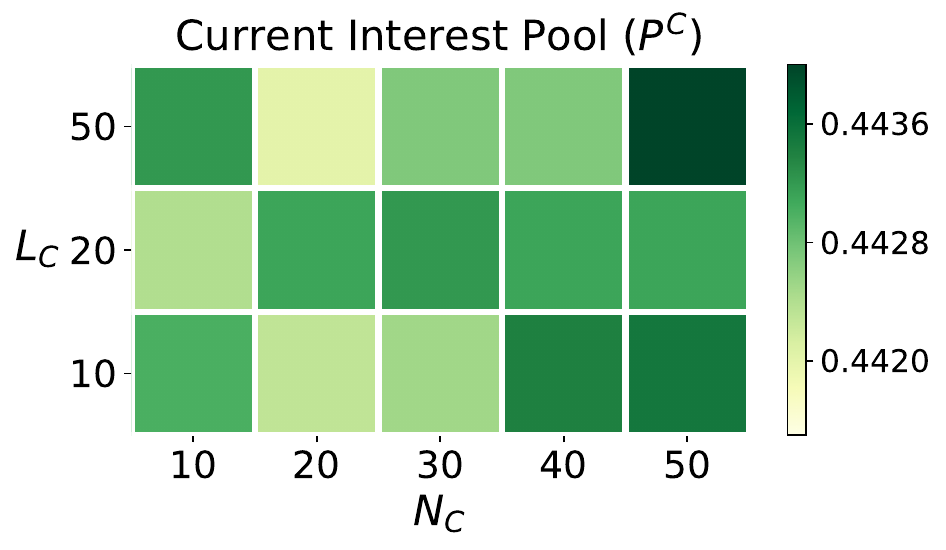}
  \caption{Impact of the number and length of interests.} 
  \label{fig:_interest_pool}
  \vspace{-0.3cm}
\end{figure}

\begin{figure}[t]
  \includegraphics[width=0.4\columnwidth]{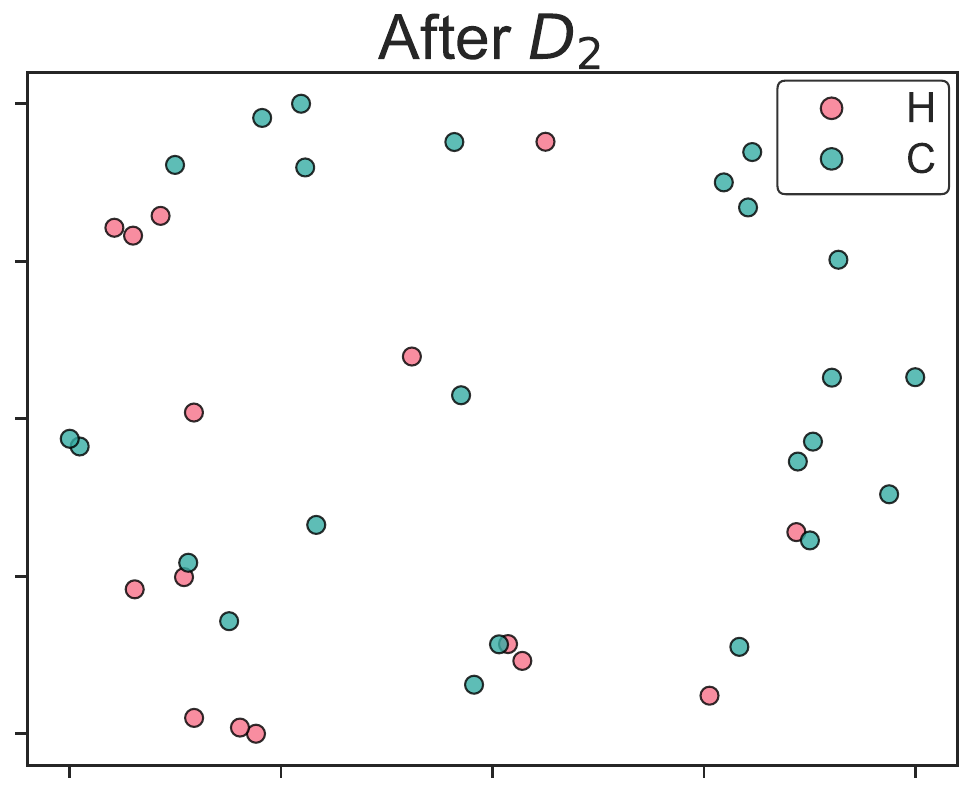}
  \includegraphics[width=0.4\columnwidth]{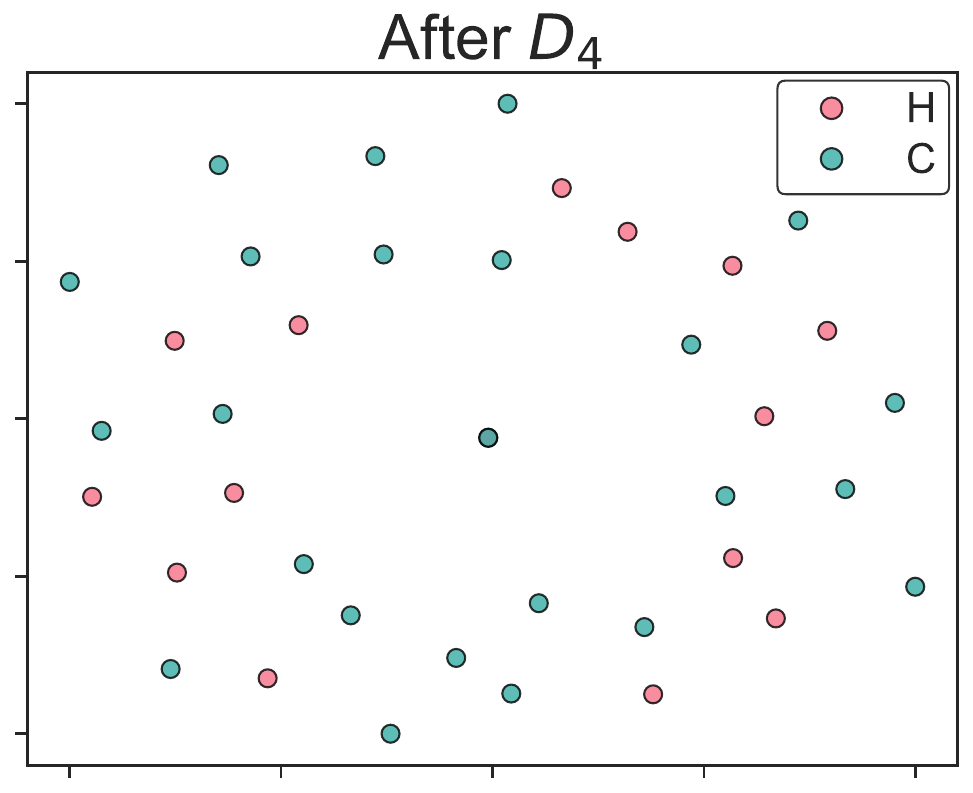}
  \caption{t-SNE visualization of interest pools.} 
  \label{fig:tsne_quality}
  \vspace{0cm}
\end{figure}

\vspace{-0.15cm}
\subsubsection{\textbf{Interest pool analysis.}}
\label{subsub:interest_pool_analysis}
%We provide an analysis to understand the behavior of interest pools.
%First, Table~\ref{tab:corr} presents the mean and variance of correlations for both historical ($H$) and current ($C$) interests across data blocks.
%The mean correlation values are close to zero with low variance across data blocks for both types of interests, indicating that each interest indeed captures mutually distinct knowledge.
Figure~\ref{fig:tsne_quality} presents the t-SNE visualization results of historical and current interest pools for data blocks \(D_2\) and \(D_4\).
We conduct a visualization of interests averaged with respect to their lengths (i.e., $L_H$ and $L_C$).
% Here, we average the lengths $L_H$ and $L_C$ of H and C interests, respectively, and then apply to t-SNE. 
We observe that as data progresses across blocks, each interest becomes more evenly distributed, indicating a progressive capture of more distinct knowledge.
By capturing this diverse knowledge, CIE can provide tailored guidance for each user, compensating for forgotten knowledge and supplementing insufficient information in incoming sequences.

\section{Conclusion}
%\vspace{-0.05cm}
\label{sec:conclusion}
We introduce \proposed, a transformer-based SR model capable of handling non-stationary data streams, which has been underexplored in previous literature.
% 제안
Building on the advantages of linear attention, we propose CSA with two novel components, CSN and CIE, enabling \proposed to both retain historical knowledge and acquire current one over time.
We also propose a pseudo-historical knowledge assignment strategy for new users to facilitate their adaptation.
% 실험
% Our experiments demonstrate the effectiveness of \proposed in terms of both knowledge acquisition and retention across data streams.
% 기대하는 바
Our experiments show that \proposed effectively captures the trajectory of user interests.
We expect that \proposed broadens the applicability of SR models to continuously changing environments.

% \section*{GenAI Usage Disclosure}
% We used ChatGPT (GPT-4o) for proofreading and polishing the English language of this manuscript. 
% All AI-generated suggestions were manually reviewed and approved by the authors. 
% No other generative AI tools were used for experimentation, data analysis, or content generation.

\section*{Ethical Considerations}
We believe this work does not pose significant negative societal impacts. Specifically, it does not raise any additional fairness, privacy, security, or safety concerns beyond standard practice.

\section*{Acknowledgement}
This work was supported by ICT Creative Consilience Program through the IITP grant funded by the MSIT (IITP-2025-RS-2020-II201819) and Basic Science Research Program through the NRF funded by the Ministry of Education (NRF-2021R1A6A1A03045425), and 2025 High-Performance Computing Support Program.

%\pagebreak
%\newpage

% \clearpage

% 참고문헌 삽입
\bibliographystyle{ACM-Reference-Format}
\balance
\bibliography{acmart}

% float 정리 후 섹션 시작
% \pagebreak
% \newpage
% \clearpage

% appendix
% \appendix

% \nobalance
% \input{sections/080Appendix}

\end{document}